\documentclass[12pt]{article}
\pdfoutput=1

\usepackage[bulletsep]{collref}
\usepackage{amssymb,graphicx}
\usepackage[intlimits]{amsmath}
\usepackage{bbm}
\usepackage[small]{subfigure}

\usepackage{MnSymbol}

\makeatletter \@addtoreset{equation}{section} \makeatother

\makeatletter
\let\old@startsection=\@startsection
\let\oldl@section=\l@section
\renewcommand{\@startsection}[6]{\old@startsection{#1}{#2}{#3}{#4}{#5}{#6\mathversion{bold}}}
\renewcommand{\l@section}[2]{\oldl@section{\mathversion{bold}#1}{#2}}
\makeatother

\makeatletter
\let\old@makecaption=\@makecaption
\def\@makecaption{\small\old@makecaption}
\makeatother

\renewcommand{\geq}{\geqslant}

\def\KK{{\cal K}}

\begin{document}

%%%%%%%%%%%%%%%%%%%%%%%%%%%%%%%%%%%%%%%%%%%%%%%%%%%%%%%%%%%%%%%%%%%%%%%%%%%%%%%%

\begin{flushright}\footnotesize
%\texttt{ITEP-TH-nn/yy}\\
\texttt{NORDITA-2014-101} \\
\texttt{UUITP-10/14}
\vspace{0.6cm}
\end{flushright}

\renewcommand{\thefootnote}{\fnsymbol{footnote}}
\setcounter{footnote}{0}

\begin{center}
{\Large\textbf{\mathversion{bold} $N=2^*$ Super-Yang-Mills Theory  \\
 at Strong Coupling}
\par}

\vspace{0.8cm}

\textrm{Xinyi Chen-Lin$^{1,2}$, James Gordon$^{1,2,3}$ and
Konstantin~Zarembo$^{1,2}$\footnote{Also at ITEP, Moscow, Russia}}
\vspace{4mm}

\textit{${}^1$Nordita, KTH Royal Institute of Technology and Stockholm University,
Roslagstullsbacken 23, SE-106 91 Stockholm, Sweden}\\
\textit{${}^2$Department of Physics and Astronomy, Uppsala University\\
SE-751 08 Uppsala, Sweden}\\
\textit{${}^3$Department of Physics and
Astronomy, University of British Columbia, 6224 Agricultural Road,
Vancouver, British Columbia V6T 1Z1}\\

\vspace{0.2cm}
\texttt{xinyic@nordita.org, jbgordon@phas.ubc.ca, zarembo@nordita.org}
%\vspace{3mm}

\vspace{3mm}

%%%%%%%%

\par\vspace{1cm}

\textbf{Abstract} \vspace{3mm}

\begin{minipage}{13cm}
 The planar $\mathcal{N}=2^*$ Super-Yang-Mills (SYM) theory is solved at large 't~Hooft coupling using localization on $S^4$. The solution permits detailed investigation of the resonance phenomena responsible for quantum phase transitions in infinite volume, and leads to quantitative predictions for the semiclassical string dual of the $\mathcal{N}=2^*$ theory.
\end{minipage}

\end{center}

\vspace{0.5cm}

%%%%%%%%%%%%%%%%%%%%%%%%%%%%%%%%%%%%%%%%%%%%%%%%%%%%%%%%%%%%%%%%%%%%%%%%%%%%%%%%

\newpage
\setcounter{page}{1}
\renewcommand{\thefootnote}{\arabic{footnote}}
\setcounter{footnote}{0}

%%%%%%%%%%%%%%%%%%%%%%%%%%%%%%%%%%%%%%%%%%%%%%%%%%%%%%%%%%%%%%%%%%%%%%%%%%%%%%%%
%%%%%%%%%%%%%%%%%%%%%%%%%%%%%%%%%%%%%%%%%%%%%%%%%%%%%%%%%%%%%%%%%%%%%%%%%%%%%%%%
\section{Introduction}

Supersymmetric localization is a way to compute path integrals in interacting field theories directly, without making any approximations \cite{Witten:1988ze}.  Our work exploits localization on $S^4$, in which case the field-theory path integral reduces to a finite-dimensional matrix model  \cite{Pestun:2007rz}.
An interesting regime that can then be explored in detail is the planar, large-$N$ limit. The strong-coupling behavior of a planar theory is generally believed to have a simple (weakly-coupled) string description. Localization  can be used to  test gauge/string duality in a very precise way, while also giving us insight into possible dynamical effects in strongly coupled gauge theories.

Localization, as a method, has obvious limitations as it requires a sufficient amount of supersymmetry. Gauge/string duality is believed to have a much broader scope, but it too is formulated precisely only in  a limited number of cases.
The model studied in this paper, $\mathcal{N}=2^*$ super-Yang-Mills theory, is special in this respect. Its partition function on $S^4$ and some select observables are calculable by localization \cite{Pestun:2007rz}, and at the same time it has a well-defined holographic dual -- the type IIB string theory on the Pilch-Warner background \cite{Pilch:2000ue}.  

The localized partition function of the $\mathcal{N}=2^*$ SYM on $S^4$ is a matrix model which at large-$N$ can be studied by standard methods  \cite{Brezin:1977sv} of random matrix theory \cite{Russo:2012kj,Buchel:2013id,Russo:2013qaa,Russo:2013kea,Russo:2013sba}. A number of observables computed with the help of localization  can be successfully compared to string-theory predictions at strong coupling. These include Wilson loops for asymptotically large contours \cite{Buchel:2013id}, and the free energy on $S^4$ \cite{Bobev:2013cja}. 

Away from the strict strong-coupling limit, localization leads to somewhat unexpected results. It turns out that the planar $\mathcal{N}=2^*$ SYM has a very complicated phase structure, undergoing an infinite number of quantum phase transitions as the 't~Hooft coupling changes from zero to infinity \cite{Russo:2013qaa,Russo:2013kea}. While these are a common phenomenon in large-$N$ theories \cite{Gross:1980he,Wadia:2012fr}, the behavior in $\mathcal{N}=2^*$ SYM is unique and differs in many respects from phase transitions in ordinary matrix models \cite{Russo:2013kea}. Similar phase transitions have also been found in QCD-like vector models \cite{Russo:2013kea,Barranco:2014tla} as well as in three \cite{Barranco:2014tla,Anderson:2014hxa,Russo:2014bda} and five  \cite{Minahan:2014hwa} dimensional theories.

\begin{figure}[t]
\begin{center}
 \centerline{\includegraphics[width=12cm]{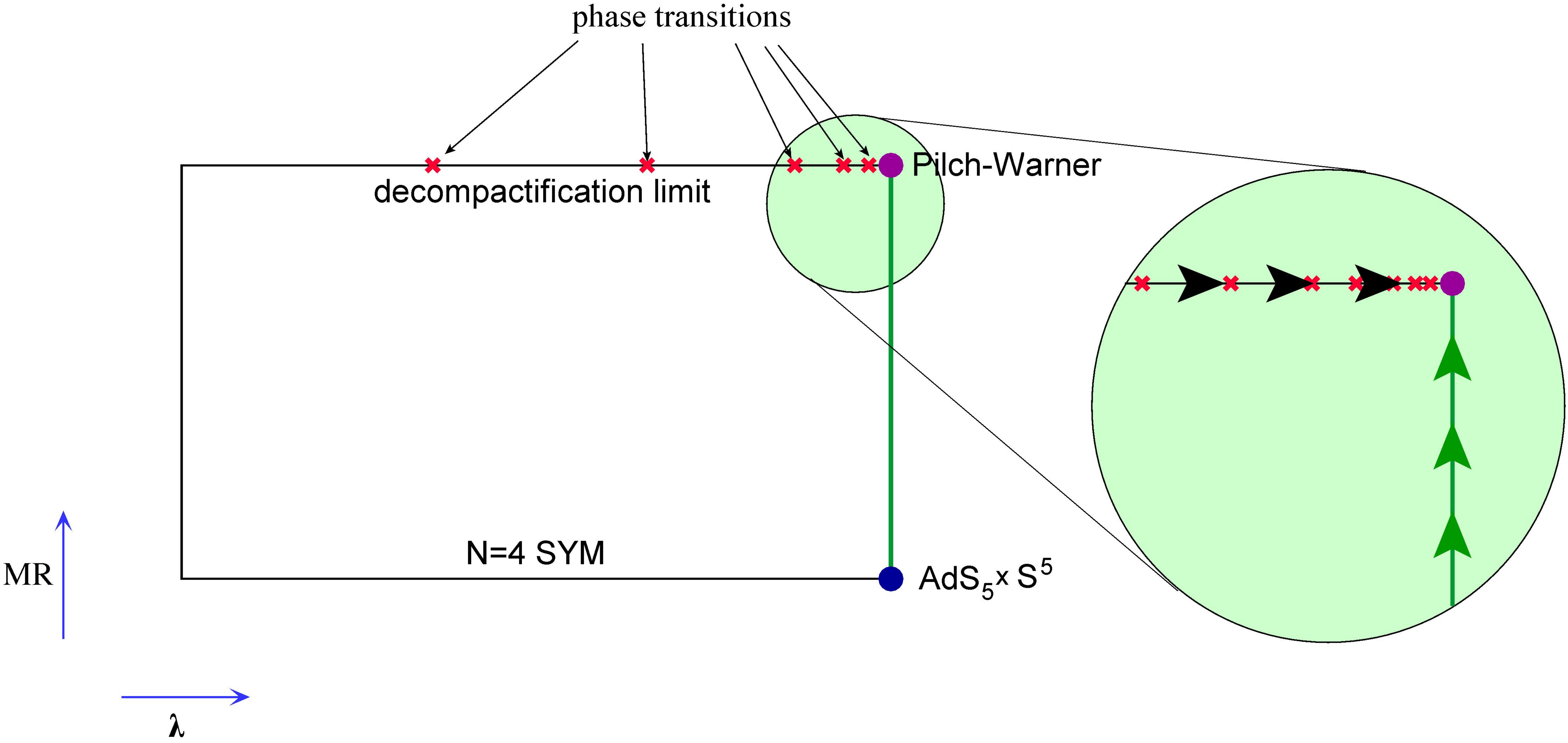}}
\caption{\label{PhD}\small The phase diagram of the planar $\mathcal{N}=2^*$ theory \cite{Russo:2013kea}. If the strong coupling limit is approached at strictly infinite radius (on $\mathbbm{R}^4$), as shown in black horizontal arrows, the theory undergoes an infinite number of phase transitions. In this paper we approach the same corner of the phase diagram along a different direction,
by varying the compactification radius while keeping the coupling  strictly infinite, as shown in green vertical arrows. Although the results may depend on the direction along which the critical point is approached, we do find the structures that cause phase transitions in the strong-coupling finite-volume solution.}
\end{center}
\end{figure}

The existence of an infinite number of phase transitions raises the question of how the non-trivial phase structure of $\mathcal{N}=2^*$ SYM is reflected in its holographic dual. Ideally, one would like to tune the 't~Hooft coupling to its critical value, but going to finite coupling is notoriously difficult in holography, as it requires quantizing string theory on a complicated supergravity background. Here we propose a different route. The transition points accumulate at infinity, and rather than varying the 't~Hooft coupling we can instead approach the accumulation point by varying the compactification radius while keeping the coupling strictly infinite (fig.~\ref{PhD}). The string dual then always remains in the classical supergravity regime. Moreover the supergravity dual for the theory on $S^4$ is explicitly known \cite{Bobev:2013cja}, so this range of parameters is potentially accessible to standard holographic calculations. 

The leading-order strong-coupling solution of the matrix model that describes $\mathcal{N}=2^*$ SYM on $S^4$ was obtained in \cite{Buchel:2013id}, and is essentially equivalent to the solution of the Gaussian matrix model. 
This result is way too simple to capture the critical behavior observed at infinite radius. 
Our aim here is two-fold. First, we would like to develop a systematic strong-coupling expansion beyond the leading order. 
Second, we want to study the approach to the critical point in the decompactification limit. The strong-coupling corrections are equivalent to quantum corrections on the string side, 
therefore, they are potentially calculable by semiclassical string quantization in the dual supergravity background. 
As we shall see, irregular structures responsible for phase transitions in the decompactification limit are already present in the first-order approximation, which opens an avenue to study the critical behavior of $\mathcal{N}=2^*$ SYM within  semiclassical string theory. 

\section{Localization}\label{sec2}

The $\mathcal{N}=2^*$ theory is the unique massive deformation of $\mathcal{N}=4$ SYM that preserves half of the rigid supersymmetry. The field content  consists of the gauge fields $A_\mu $, their scalar superpartners $\Phi $ and $\Phi '$, complex scalars $Z_{1,2}$ from the hypermultiplet, and the fermions. All the fields are in the adjoint of the gauge group which we take to be $SU(N)$. The Lagrangian of the $\mathcal{N}=2^*$ theory is obtained from that of $\mathcal{N}=4$ SYM by giving common mass $M$ to the hypermultiplet fields and adding certain Yukawa couplings necessary for supersymmetry.  

The scalars in the vector multiplet can condense along the flat directions of the potential 
$V\sim \mathop{\mathrm{tr}}[\Phi ,\Phi ']^2$:
\begin{equation}\label{Phi}
 \left\langle \Phi \right\rangle=\mathop{\mathrm{diag}}\left(a_1,\ldots ,a_N\right),
\end{equation}
thus breaking the gauge symmetry to $U(1)^{N-1}$. The $ij$ components of the vector-multiplet fields then acquire masses $m_{ij}^{v}=|a_i-a_j|$ while hypermultiplet masses are $m_{ij}^h=|a_i-a_j\pm M|$. The vector multiplets get light as soon as the eigenvalues $a_i$ and $a_j$ approach one another -- the diagonal states with $i=j$ are photons of the unbroken $U(1)^{N-1}$. In the matrix model these massless states do not lead to any dramatic effects, because the contribution of the light fields is counteracted by the Vandermonde repulsion of the matrix eigenvalues. The effects due to light hypermultiplets are much more dramatic. Massless hypermultiplets appear once the distance between two eigenvalues gets close to $M$. This resonance phenomenon plays an important r\^ole in shaping the phase structure of the $\mathcal{N}=2^*$ SYM in the large-$N$ limit.

An important characteristic of the theory at large $N$ is the master field, characterized by the eigenvalue density:
\begin{equation}\label{master-field}
 \rho (x)=\left\langle \frac{1}{N}\sum_{i=1}^{N}\delta \left(x-a_i\right)\right\rangle.
\end{equation}
Our goal is to study the exact master field of the $\mathcal{N}=2^*$ theory on a four-sphere of radius $R$ at large 't~Hooft coupling $\lambda \equiv g_{\rm YM}^2N\gg 1$. 
Up to some point, we will keep the full dependence on the dimensionless parameter $MR$, 
and will then separately study the decompactification limit $R\rightarrow \infty $, where most of the interesting phenomena occur. 
The resulting theory can be viewed as $\mathcal{N}=2^*$ SYM in flat space in a particular vacuum state selected by compactification. The same vacuum is singled out by conformal perturbation theory in $\mathcal{N}=4$ SYM, and also by AdS/CFT duality for the Pilch-Warner background (see \cite{Russo:2013sba} for a more detailed discussion of the vacuum selection in  this context).

 Supersymmetric localization reduces the path integral of $\mathcal{N}=2^*$ SYM on $S^4$ to an $(N-1)$-dimensional eigenvalue integral \cite{Pestun:2007rz}:
\begin{equation}
\label{mint}
Z=\int d^{N-1} a\, 
\prod _{i<j}\frac{(a_i-a_j)^2H^2(a_i-a_j)}{H(a_i-a_j-M)H(a_i-a_j+M)}
\,{\rm e}\,^{-\frac{8\pi^2N}{\lambda}\sum_i a_i^2 } \left|\mathcal{Z}_{\rm inst}\right|^2,
\end{equation}
where
\begin{equation}\label{functionH}
 H(x)\equiv \prod_{n=1}^\infty \left(1+\frac{x^2}{n^2}\right)^n \,{\rm e}\,^{-\frac{x^2}{n}} .
\end{equation}
We will set the instanton contribution to zero, $\mathcal{Z}_{\rm inst}=1$, because at large $N$ instantons are exponentially suppressed.

Localization allows one to compute some special correlation functions, for example the expectation value of the Wilson loop for the big circle of $S^4$. The localizable loop operator couples to the scalar $\Phi $ of the vector multiplet, in addition to the usual path-ordered vector coupling:
\begin{equation}\label{Wilson}
 W(C)=\left\langle \frac{1}{N}\,\mathop{\mathrm{tr}}{\rm P}\exp
 \oint_{C}ds\,\left(iA_\mu \dot{x}^\mu +\Phi |\dot{x}|\right)
 \right\rangle.
\end{equation}
If $C$ is the equatorial circle of $S^4$, the Wilson loop can be computed by just substituting the constant classical value (\ref{Phi}) for $\Phi $ and averaging over the eigenvalues with the weight given by the partition function of the  matrix model (\ref{mint}):
\begin{equation}\label{W(C)}
 W(C)=\left\langle \frac{1}{N}\sum_{i}^{}\,{\rm e}\,^{2\pi a_i}\right\rangle
 =\int_{-\mu }^{\mu }dx\,\rho (x)\,{\rm e}\,^{2\pi x}.
\end{equation}

The eigenvalue integral (\ref{mint}) is of the saddle-point type at large $N$, and the partition function is dominated by a single equilibrium configuration of $a_i$'s when $N\rightarrow \infty $. 
The equilibrium condition can be written as a singular integral equation for the master field (\ref{master-field}):
\begin{equation}
 \label{inteq}
\strokedint_{-\mu}^\mu dy\, \rho(y)S(x-y)= \frac{8\pi^2}{\lambda}\ x,
\end{equation}
where the kernel is given by
\begin{equation} \label{kernel}
 S(x)= \frac{1}{x }-\KK(x)+\frac{1}{2}\,\KK(x +M)+\frac{1}{2}\,\KK(x -M).
\end{equation}
The function $\KK(x)$ is the logarithmic derivative of $H(x)$:
\begin{equation}\label{KK-def}
 \KK(x)=-\frac{H'(x)}{H(x)}=2x\sum_{n=1}^\infty \left(\frac{1}{n} -\frac{n}{n^2+x^2}\right)
 =x\left(\psi \left(1+ix\right)+\psi \left(1-ix\right)-2\psi (1)\right).
\end{equation}
The equations above are written in units in which the radius of the sphere is set to one. The dependence on $R$ can be recovered by rescaling $M\rightarrow MR$, $x\rightarrow xR$ and so on. We will keep the dimensionless units in this paper, but it is important to remember that the decompactification limit $R\rightarrow \infty $ is now traded for the infinite mass limit $M\rightarrow \infty $.

The solution  of the saddle-point equations at infinitely strong coupling was found in \cite{Buchel:2013id}, and is very simple. Let us repeat the derivation here. The restoring force on the right hand side of  (\ref{inteq}) is very small at large $\lambda $ and the eigenvalues spread over wider and wider intervals as $\lambda $ grows. It is consequently true that both $\mu \gg M$ and $\mu \gg 1$ when $\lambda\gg1 $. Since $x -y\sim \mu $, the difference operator in (\ref{kernel}) can be replaced by the second derivative, and the kernel function $\KK(x)$ can be replaced by its asymptotics at large values of the argument,
\begin{equation}\label{largex-KK}
 \KK(x)\simeq x\ln x^2\qquad \left(x\rightarrow \infty \right).
\end{equation}
We thus have
\begin{equation}\label{approxKK}
 \frac{1}{2}\,\KK(x+M)+\frac{1}{2}\,\KK(x-M)- \KK(x)
 \approx \frac{1}{2}\,M^2\KK''(x)\approx \frac{M^2}{x}\,.
\end{equation}
The saddle-point equation reduces to that of the Gaussian matrix model:
\begin{equation}\label{LOeq}
 \strokedint_{-\mu}^\mu dy\, \rho (y)\,\frac{1+M^2}{x-y}=\frac{8\pi^2}{\lambda}\ x,
\end{equation}
whose solution is Wigner's semicircle law:
\begin{equation}\label{Wign-scir}
 \rho _\infty (x)=\frac{2}{\pi \mu ^2}\,\sqrt{\mu ^2-x^2}
\end{equation}
with
\begin{equation}\label{leading-mu}
 \mu =\frac{\sqrt{\lambda \left(1+M^2\right)}}{2\pi }\,.
\end{equation}
Taking the decompactification limit $M\rightarrow \infty $ leads to the Wigner distribution of width $\sqrt{\lambda }M/2\pi $, in precise agreement with the D-brane probe analysis of the Pilch-Warner solution \cite{Buchel:2000cn}. The same result can be also derived within the Seiberg-Witten theory \cite{Billo:2014bja}.

From this solution we can calculate the expectation value of the Wilson loop (\ref{W(C)}):
\begin{equation}\label{LoopofWilson}
 W(C)\simeq \,{\rm const}\,\lambda ^{-\frac{3}{4}}\,{\rm e}\,^{\sqrt{\lambda \left(1+M^2\right)}}.
\end{equation}
The leading exponential behavior corresponds to the minimal area law in the dual gravitational description, and at $M\rightarrow \infty $ one finds a precise match with the area law in the Pilch-Warner background \cite{Buchel:2013id}. The prefactor corresponds to the contribution of string fluctuations, potentially calculable by semiclassical string quantization. One of our goals is to compute this prefactor from the matrix model.

Using the leading-order solution (\ref{Wign-scir}), we would get $\sqrt{2/\pi }(1+M^2)^{-3/4}$ for the constant of proportionality in (\ref{LoopofWilson}). We do not display this result in the equation, because it is actually incorrect. Indeed, the weight in the integral (\ref{W(C)}) is exponentially peaked at the largest eigenvalue, and the most important contribution comes from $x$ very close to $\mu $, namely from $\mu-x \sim 1$. The eigenvalue density there is very small, so that corrections to the leading-order Wigner distribution (\ref{Wign-scir}) are  $O(1)$ and not $O(1/\sqrt{\lambda })$ as one might naively expect. Thus, to compute the Wilson loop, we need to know the endpoint behavior of the eigenvalue distribution exactly. 

The importance of the endpoint region can be understood from a different perspective. Suppose that we want to calculate the next correction  in $1/\sqrt{\lambda }$ to the eigenvalue density. How do we proceed? The first idea that comes to mind is to expand (\ref{approxKK}) to higher orders in $1/x$; however, careful inspection of the resulting equations shows that this idea does not work. In fact, the equation stays the same at the next order in $1/\sqrt{\lambda }$. What changes is the boundary behavior of the density. The general solution to (\ref{LOeq}) reads
\begin{equation}\label{corrdens}
 \rho (x)=\frac{8\pi }{\lambda \left(1+M^2\right)}\,\sqrt{\mu ^2-x^2}
 +\frac{\beta }{\mu \sqrt{\mu ^2-x^2}}\,.
\end{equation}
The second term, which we normalized to obtain a $1/\sqrt{\lambda }$ correction to the Wigner distribution, satisfies the homogeneous form of the integral equation (\ref{LOeq}). The coefficient $\beta$ is thus  not fixed by the equations; nor is it fixed by the normalization condition of the density, which is supposed to determine the endpoint position $\mu $.

This freedom to choose any coefficient for the second term may look worrisome. Moreover, the density vanishes at the endpoints for any finite $\lambda $, while the solution above blows up at the edges of the interval. A resolution of these apparent contradictions lies in the fact that the second term in (\ref{corrdens}) cannot be treated as a small correction near the endpoints. The two terms in (\ref{corrdens}) become comparable at  $\mu-x\sim 1 $, where the naive strong-coupling expansion breaks down.
 We shall see that the most interesting phenomena, responsible for the phase transitions in infinite volume, happen precisely in this regime. We will eventually fix the constant $\beta $ and then $\mu $ by matching the exact solution in the near-endpoint region to the asymptotic solution (\ref{corrdens}) in the bulk of the eigenvalue distribution.

\section{Solution at strong coupling}

\subsection{Wiener-Hopf problem}

As we concluded above, the strong-coupling expansion breaks down near the endpoints of the eigenvalue distribution. The integral equation, therefore, has to be analyzed separately in this regime. We recall that the support of the eigenvalue density becomes very large, $\mu \gg 1$,  at strong coupling. We are interested in the behavior at $x$ close to $\mu $, with $\mu -x\sim 1$.
To gain some intuition we can start with the leading-order solution  (\ref{Wign-scir}). Introducing the variable
\begin{equation}\label{xi}
 \xi =\mu -x,
\end{equation}
  we find at $\xi \sim 1$:
\begin{equation}\label{xi-Wigner}
 \rho _\infty (x)\simeq \frac{2^{\frac{3}{2}}}{\pi \mu ^{\frac{3}{2}}}\,\sqrt{\xi }.
\end{equation}
We thus expect that the exact density near the upper endpoint has the form
\begin{equation}\label{fofxi-def}
 \rho (x)=\frac{2^{\frac{3}{2}}}{\pi \mu ^{\frac{3}{2}}}\,f(\xi ),
\end{equation}
where  $f(\xi )$ is a scaling function that does not depend on $\lambda $.

Since $\rho _\infty $ is a good approximation in the bulk of the eigenvalue distribution, (\ref{fofxi-def}) should approach (\ref{xi-Wigner}) away from the endpoint. This fixes the boundary conditions on $f(\xi )$
 at large $\xi $:
\begin{equation}\label{b.c.}
 f(\xi )\simeq \sqrt{\xi }\qquad \left(\xi \rightarrow \infty \right).
\end{equation}
At $\xi \sim 1$, $f(\xi )$ deviates from $\sqrt{\xi }$ by an $O(1)$ amount. 

An integral equation for $f(\xi )$ can be obtained by the following trick: the exact saddle-point equation (\ref{inteq}) can be re-written as
\begin{equation}
\strokedint_{-\mu}^\mu dy\, \left[\rho(y) S(x-y)-\rho_\infty (y) \,\frac{1+M^2}{x-y}\right]=0,
\end{equation}
where we have made use of  (\ref{LOeq}). This step entails no approximations. We can now notice that for $x$ close to $\mu $ the weight in the $y$ integral is peaked near the endpoint. We can thus introduce the scaling variable (\ref{xi}), replace $\rho (y)$ and $\rho _\infty (y)$  by their scaling forms (\ref{fofxi-def}), (\ref{xi-Wigner}),
and extend the limit of integration over the scaling variable to infinity:
\begin{equation}\label{WienerHopf}
 \strokedint_{0}^\infty  d\eta \,\left[ f(\eta ) S(\eta -\xi)-\frac{(1+M^2)\sqrt{\eta }}{\eta -\xi }\right]=0.
\end{equation}
The  integral over $\eta $ converges at the upper limit due to the asymptotic form of $K(x)$ given in (\ref{largex-KK}), (\ref{approxKK})  and the boundary condition (\ref{b.c.}). 

The integral equation (\ref{WienerHopf}) is of the Wiener-Hopf type. It can be brought to the standard Wiener-Hopf form by introducing the regularized scaling function with a better behavior at infinity:
\begin{equation}\label{gofxi-def}
 g(\xi )=f(\xi )-\sqrt{\xi }.
\end{equation}
This function satisfies the equation
\begin{equation}\label{main}
 \strokedint_0^\infty d\eta \,g(\eta )S(\eta -\xi)=F(\xi ),
\end{equation}
where
\begin{equation}\label{F-def}
 F(\xi )= \strokedint_0^\infty d\eta \,\sqrt{\eta } \: \left[ \frac{1+M^2}{\eta -\xi }-S(\eta -\xi) \right].
\end{equation}
Both integrals here converge -- in the first case because $g(\eta  )\sim 1/\sqrt{\eta  }$ at large $\eta  $,
and in the second case because the kernel behaves as $1/(\eta -\xi )^3$ at large $\eta $, as a consequence of \eqref{approxKK}.

\subsection{Exact solution}

In order to solve the integral equation \eqref{main}, 
let us first assume that the function $g(\xi )$ is defined on the whole real axis, but is equal to zero for $\xi <0$. 
Then, the integral on the left-hand side of (\ref{main}) takes the convolution form:
\begin{equation}
 S*g(\xi )=F(\xi )+\theta (-\xi )X(\xi ),
\end{equation}
where $X(\xi )$ is an arbitrary function whose appearance reflects the fact that the original equation only holds for $\xi >0$. 

After the equation is written in the convolution form, it can be brought to an algebraic form by Fourier transform:
\begin{equation}
 g(\xi )=\int_{-\infty }^{+\infty }\frac{d\omega }{2\pi }\,\,{\rm e}\,^{-i\omega \xi }\hat{g}(\omega ).
 \label{FTgxi}
\end{equation}
After the Fourier transform we get:
\begin{equation}\label{WH-equation}
 \hat{S}(\omega )\hat{g}(\omega )=\hat{F}(\omega )+X_-(\omega ),
\end{equation}
where $X_-(\omega )$ is a function that has no singularities in the lower half-plane of complex $\omega $, because its Fourier image vanishes on the positive real semi-axis. 
The subindex $+$ will be used similarly, but to indicate analyticity in the upper half-plane.
In fact, $\hat{g}(\omega )=\hat{g}_+(\omega)$.
The explicit expressions for the remaining terms are:
\begin{eqnarray}
 \hat{S}(\omega )&=&i\pi \mathop{\mathrm{sign}}\omega \left(1+
 \frac{\sin^2\frac{M\omega }{2}}{\sinh^2\frac{\omega }{2}}
 \right)
 \\
 \hat{F}(\omega )&=&i\pi \mathop{\mathrm{sign}}\omega \left(M^2-
 \frac{\sin^2\frac{M\omega }{2}}{\sinh^2\frac{\omega }{2}}
 \right)\frac{i^{\frac{3}{2}}\sqrt{\pi }}{2\omega \sqrt{\omega +i\epsilon }}\,,
\end{eqnarray}
which can be obtained from the definitions (\ref{kernel}), (\ref{KK-def}) and (\ref{F-def}), as well as the following representation of $\KK''(x)$:
\begin{equation}
	\KK''(x) = 4 \int_0^{\infty}\! d\omega \, \frac{\omega^2 \sin 2x\omega}{\sinh^2 \omega} .
\end{equation}
Here, the sign function should be understood in terms of analytic regularization:
\begin{equation}\label{sign-reg}
 \mathop{\mathrm{sign}}\omega =\lim_{\epsilon \rightarrow 0}\frac{\sqrt{\omega +i\epsilon }}{\sqrt{\omega -i\epsilon }}\,,
\end{equation}
where the square root $\sqrt{\omega \pm i\epsilon }$ has a branch cut extending from $\mp i\epsilon $ up to infinity along the negative/positive imaginary semi-axis.

The solution of the Wiener-Hopf problem (\ref{WH-equation}) is based on the factorization formula
\begin{equation}\label{factorization}
 \hat{S}(\omega )=\frac{1}{G_+(\omega )G_-(\omega )}.
\end{equation}
In our case, the functions $G_\pm$ have the form:
\begin{equation}\label{Gplusminus}
 G_\pm(\omega )=\left(\frac{M^2+1}{i\pi }\right)^{\frac{1}{2}}\left(\omega \pm i\epsilon \right)^{\mp\frac{1}{2}}\,{\rm e}\,^{\mp \frac{ i\phi \omega}{2\pi } }\,
 \frac{\Gamma \left(\frac{M\mp i}{2\pi }\,\omega \right)
 \Gamma \left(-\frac{M\pm i}{2\pi }\,\omega \right)}{\Gamma ^2\left(\mp\frac{i\omega }{2\pi }\right)}\,,
\end{equation}
where
\begin{equation}
 \phi =2M\arctan M-\ln\left(M^2+1\right).
\end{equation}
These formulae follow from factorization of trigonometric and hyperbolic functions in terms of the gamma function:
\begin{equation}\label{gamma-gamma}
 \Gamma(x)\Gamma(1-x)=\dfrac{\pi}{\sin\pi x }
\end{equation}
The phase term $ \,{\rm e}\,^{\mp \frac{ i\phi \omega}{2\pi } }$ is introduced to cancel the bad asymptotics of the combination of gamma functions at large imaginary $\omega$, which can be inferred from the Stirling formula. Without these factors the functions $G_\pm (\omega )$ would exponentially grow in their respective domains of analyticity.

\begin{figure}[t]
\begin{center}
 \centerline{\includegraphics[width=9cm]{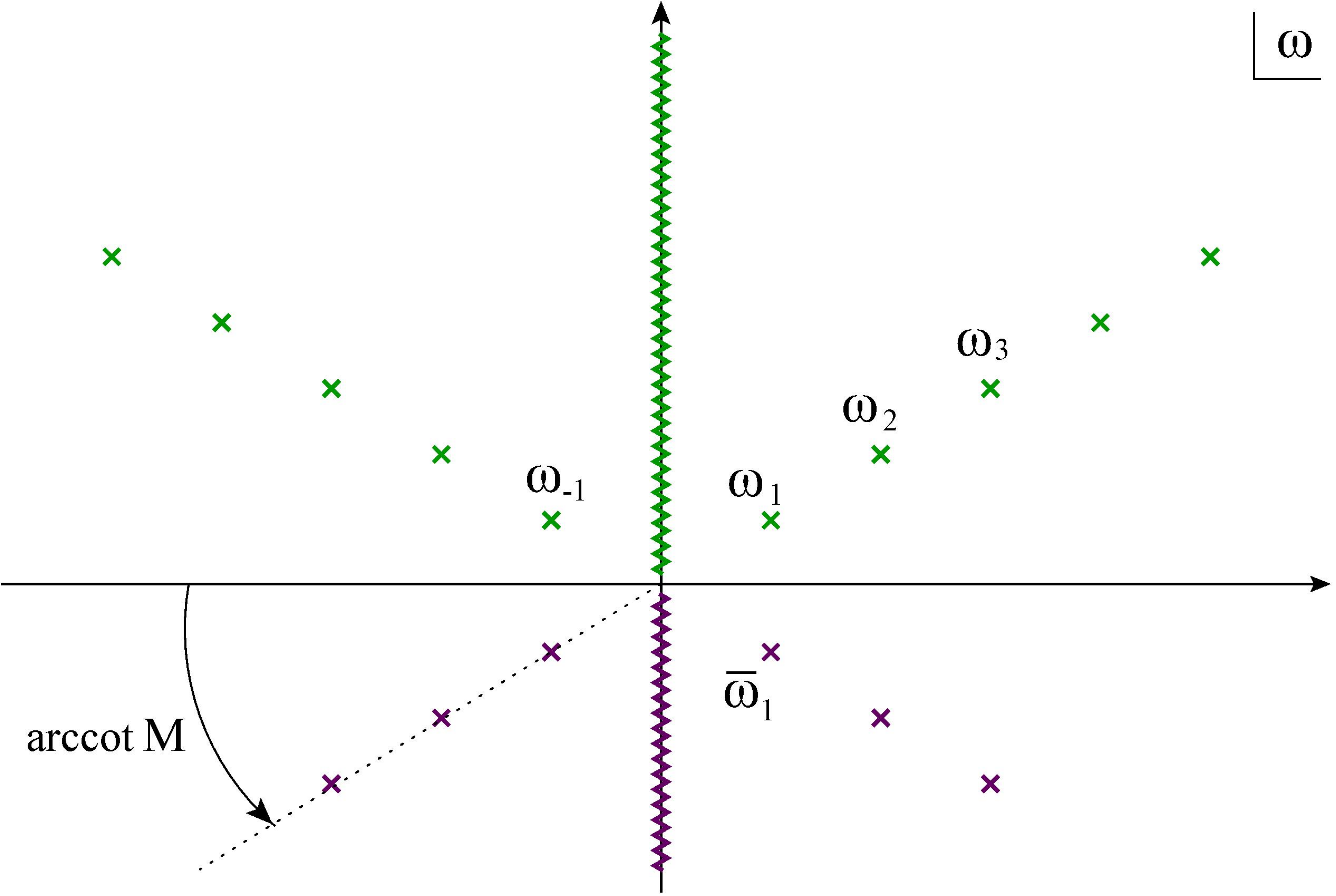}}
\caption{\label{a-structureofG}\small The singularities of the inverse kernel, $\hat{S}^{-1}(\omega )$, are two cuts along positive and negative imaginary semi-axes, and simple poles at $\omega =\omega _n$ and $\omega =\bar{\omega }_n$. The functions $G_-$ and $G_+$ inherit singularities in respectively the upper and lower half-planes.}
\end{center}
\end{figure}

The inverse kernel of the integral equation $\hat{S}^{-1}(\omega )$ has a cut along the imaginary axis, due to  the sign function, which we break in two parts by regularization (\ref{sign-reg}). It also has simple poles at $\omega =\omega _n$ and $\omega =\bar{\omega }_n$,   $n\neq 0$, where
\begin{equation}\label{poles}
 \omega_n=\frac{2\pi \left(Mn+i|n|\right)}{M^2+1}\,.
\end{equation}
The factorization (\ref{factorization}) assigns the cut in the upper half-plane and the poles at $\omega =\omega _n$ to $G_-$, while $G_+$ has a cut in the lower half-plane and poles at $\omega =\bar{\omega }_n$  (fig.~\ref{a-structureofG}). The poles of $G_\pm$ lie on the straight lines that make an angle  $\alpha =\mathop{\mathrm{arccot}} M$ with the real axis. After the Fourier transform back to $\xi $ space, the poles will create resonances. For generic $M$ the resonances are damped because $\mathop{\mathrm{Im}}\omega _n\sim \mathop{\mathrm{Re}}\omega _n$, but
when $M$ becomes large, the poles pinch the real axis and cause oscillations in the $\xi $ space, with periods that are integer multiples of $M$. This behavior is a manifestation of the nearly massless hypermultiplets that we have discussed in sec.~\ref{sec2}. We shall study the large-$M$ behavior of the solution in much detail in the next section.

Returning to the Wiener-Hopf equation, we can use factorization  of the kernel to rewrite  \eqref{WH-equation} as
\begin{equation}\label{last-eq}
\dfrac{\hat{g}_+(\omega )}{G_+(\omega)}=G_-(\omega)\hat{F}(\omega )+G_-(\omega)X_-(\omega ).
\end{equation}
Defining the projection on the positive/negative-frequency part of a function via contour integration:
\begin{equation}\label{+projection}
 \mathcal{F}_\pm (\omega )=\pm \int_{-\infty }^{+\infty }\frac{d\omega '}{2\pi i}\,\,\frac{\mathcal{F}(\omega ')}{\omega '-\omega \mp i\epsilon }\,,
\end{equation}
we can project out the $-$ term in the last equation, and thus find the  solution to the Wiener-Hopf problem:
\begin{equation}\label{answer}
 \hat{g}_+(\omega )=G_+(\omega )\left(G_-\hat{F}\right)_+ (\omega ).
\end{equation}
It is important here that $G_+(\omega)^{-1}$ is also an analytic function in the upper half-plane of complex $\omega $, and thus the left hand side of (\ref{last-eq}) is a $+$ function. 

In order to compute the positive projection of $ (G_-\hat{F}) (\omega )$, let us discuss the analytic structure of this function first.
Since the branch cut of $G_-$ cancels in the product $G_-\hat{F}$, the latter is
a meromorphic function with simple poles at $\omega =\omega _n$, $n=\pm 1, \pm 2,\ldots $ and double poles at $\omega =-2\pi im$, $m=1,2,\ldots $. All the double poles lie in the lower half-plane, and the integral in (\ref{+projection}) can be done by closing the contour of integration
  in the upper half-plane and picking up the poles at $\omega $ and $\omega _n$:
\begin{equation}\label{expanded-solution}
\hat{g}(\omega )=\frac{\hat{F}(\omega )}{\hat{S}(\omega )}-G_+(\omega )\sum_{n\neq 0}^{}\frac{\hat{F}(\omega _n)}{\omega -\omega _n}
 \mathop{\mathrm{res}}_{z=\omega _n}G_-(z).
\end{equation}
The first term is the naive Fourier transform that would solve the integral equation on the whole real axis. Since we need a solution identically equal to zero for $\xi <0$, its Fourier transform must be analytic in the upper half plane. The r\^ole of the last term is to subtract the singularities of the first term in order to make the solution a $+$ function.  

Explicitly, we get:
\begin{align}\label{final-g}
 \hat{g}(\omega )&=\frac{i^{\frac{3}{2}}\sqrt{\pi }}{2\omega \sqrt{\omega +i\epsilon }}
 \left[
 \frac{M^2\sinh^2\frac{\omega }{2}-\sin^2\frac{M\omega }{2}}{\sinh^2\frac{\omega }{2}+\sin^2\frac{M\omega }{2}}
\right.
\nonumber \\
 &\left.
 +\left(M^2+1\right)^2\omega \,{\rm e}\,^{- \frac{i\phi \omega}{2\pi } }\,
 \frac{\Gamma \left(\frac{M- i}{2\pi }\,\omega \right)
 \Gamma \left(-\frac{M+ i}{2\pi }\,\omega \right)}{\Gamma ^2\left(-\frac{i\omega }{2\pi }\right)}
  \right.
\nonumber \\
 &\left.\times 
 \sum_{n=1}^{\infty }
 \frac{\left(-1\right)^n}{nn!}
 \left(
 \frac{\,{\rm e}\,^{\frac{i\phi n}{M-i}}}{\omega -\frac{2\pi n}{M-i}}\,\,\frac{\Gamma \left(\frac{M+i}{M-i}\,n\right)}{\Gamma ^2\left(\frac{i}{M-i}\,n\right)}
 +
 \frac{\,{\rm e}\,^{-\frac{ i\phi n}{M+i}}}{\omega +\frac{2\pi n}{M+i}}\,\,\frac{\Gamma \left(\frac{M-i}{M+i}\,n\right)}{\Gamma ^2\left(-\frac{i}{M+i}\,n\right)}
 \right)
 \right] 
\end{align}
This is our final expression which in general cannot be further simplified.

The solution in the $\xi $ space is the inverse Fourier transform of (\ref{final-g}).
Since $\hat{g}(\omega )$ has a relatively simple structure of singularities in the lower half-plane, its Fourier transform can be computed using the residue theorem, which results in a double infinite sum representation for $g(\xi )$.
The final expression \eqref{gxisolution} and the details of the derivation are given in the appendix. This expression is very convenient for numerical evaluation of the function $g(\xi )$, but many quantities of interest, such as the Wilson loop expectation value, can be calculated directly from the Fourier representation. 

The Wilson loop is an eigenvalue average with the exponential weight. Writing the defining equation \eqref{W(C)}
in terms of the endpoint variable $\xi$, and using \eqref{fofxi-def} and \eqref{gofxi-def}, we get
\begin{equation}
	W(C) = \frac{2^{3/2}}{\pi \mu^{3/2}} \, \,{\rm e}\,^{2\pi \mu}\int_0^{\infty} d\xi \:\left(g(\xi)+\sqrt{\xi}\right)\,{\rm e}\,^{-2\pi \xi},
\end{equation}
where as before, the integration domain is extended to infinity, allowed by the strong coupling limit. 
This is a Laplace integral, and the part with $g(\xi)$ is simply $\hat{g}(2\pi i)$. Therefore, the Wilson loop is:
\begin{equation}\label{WL-g}
	W(C)=\frac{2^{3/2}}{\pi \mu^{3/2}} \, \,{\rm e}\,^{2\pi \mu}\left(\hat{g}(2\pi i)+\frac{1}{2^{5/2} \pi}\right).
\end{equation}
The last term in the brackets is the contribution of the uncorrected Wigner distribution, which gives the prefactor quoted after (\ref{LoopofWilson}). The Wiener-Hopf term $\hat{g}(2\pi i)$ is the correction produced by the distortion of the eigenvalue distribution near the endpoint. We study it in more detail in the next subsection.

\subsection{General structure of solution}\label{gen-structure-sec}

\begin{figure} 
\begin{center}
\includegraphics[width=0.6\textwidth]{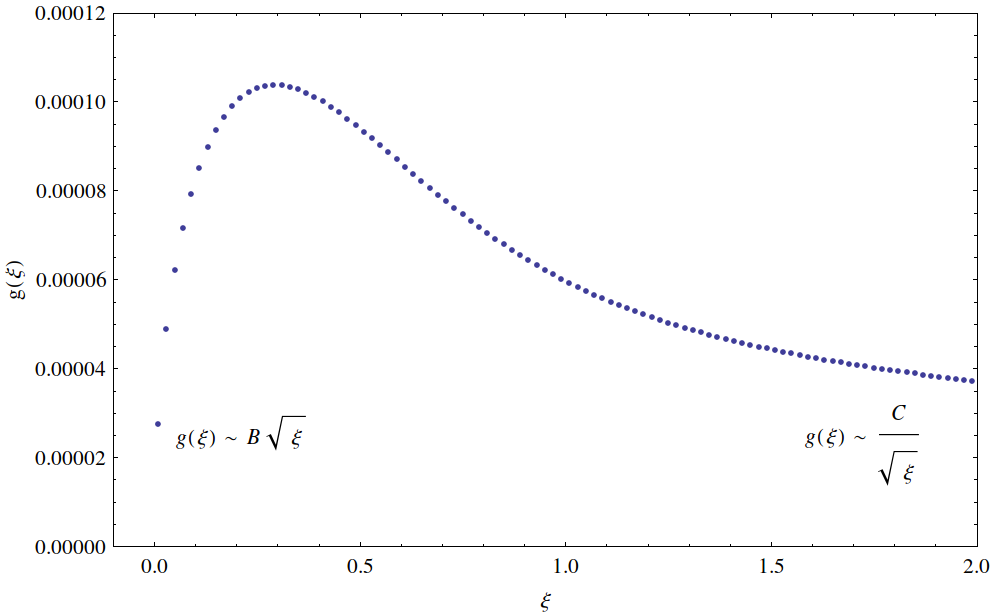} 
\end{center}
\caption{\label{gxi_smallM} $g(\xi)$ for mass $M=0.5$. }
\end{figure} 

For moderate and small values of the mass 
-- $M$ of order 1 and below --
% -- when $M$ is not too big -- 
$g(\xi )$ is a featureless function whose asymptotic behavior is prescribed by the boundary conditions of the Wiener-Hopf problem:
\begin{equation}\label{BC-def}
 g(\xi )\stackrel{\xi \rightarrow 0}{\simeq} B\sqrt{\xi },
 \qquad 
  g(\xi) \stackrel{\xi \rightarrow \infty }{\simeq}\dfrac{C}{\sqrt{\xi}},
\end{equation}
where the coefficients $B$ and $C$ depend on the mass. The plot in fig.~\ref{gxi_smallM} shows $g(\xi )$ at $M=1/2$ for illustration.

In view of (\ref{fofxi-def}), (\ref{gofxi-def}), the constant $B$ determines the endpoint behavior of the eigenvalue density:
\begin{equation}\label{nearB}
 \rho (x)=\frac{2^{\frac{3}{2}}(1+B)}{\pi \mu ^{\frac{3}{2}}}\,\sqrt{\mu -x}
 \qquad \left(x\rightarrow \mu \right).
 \end{equation}
 The constant $C$ will be later used to compute $1/\sqrt{\lambda }$ corrections in the bulk of the eigenvalue distribution. We also introduce 
\begin{equation}
 A=2^{5/2}\pi \hat{g}(2\pi i)+1,
\end{equation}
which determines the normalization factor in the expectation value of the Wilson loop:
\begin{equation}\label{W->A}
 W(C)=\frac{A}{2\pi ^2\mu ^{3/2}}\,\,{\rm e}\,^{2\pi \mu }.
\end{equation}
For pure Wigner distribution $A=1$, but $A$ is also a non-trivial function of the mass.
 
The constants $B$ and $C$ can be read off from the asymptotic behavior of $\hat{g}(\omega )$:
\begin{equation}\label{BCfromg}
 \hat{g}(\omega )\stackrel{\omega \rightarrow \infty }{\simeq }
 \frac{i^{\frac{3}{2}}\sqrt{\pi } B}{2\omega^{\frac{3}{2}}}\,,
 \qquad 
 \hat{g}(\omega) \stackrel{\omega \rightarrow 0}{\simeq }\dfrac{\sqrt{i \pi}C}{\sqrt{\omega}}\,.
\end{equation}
The constant $A$ is also expressed explicitly in terms of $\hat{g}(\omega )$. Therefore to compute these constants we do not need to perform the inverse Fourier transform back to the $\xi $ representation. Explicitly:
\begin{align}
\label{Acoeff}
A&=\frac{2\pi M\left(M^2+1\right)^2\,{\rm e}\,^{\phi}}{\sinh \pi M}
\sum_{n=1}^{\infty }
 \frac{\left(-1\right)^n}{nn!}\,
\mathop{\mathrm{Re}}\left(
 \frac{\,{\rm e}\,^{\frac{i\phi n}{M-i}}}{1+\frac{in}{M-i}}
 \,\frac{\Gamma \left(\frac{M+i}{M-i}\,n\right)}{\Gamma^2 \left(\frac{i}{M-i}\,n\right)}
 \right)
 \\
\label{Bcoeff}
 B&=M^2+2\left(M^2+1\right)^{\frac{3}{2}}\sum_{n=1}^{\infty }
 \frac{\left(-1\right)^n}{nn!}\,
\mathop{\mathrm{Re}}\left(
 \,{\rm e}\,^{\frac{i\phi n}{M-i}}
 \,\frac{\Gamma \left(\frac{M+i}{M-i}\,n\right)}{\Gamma^2 \left(\frac{i}{M-i}\,n\right)}
 \right)
 -\frac{M^2+1}{\pi }\,\arctan M
 \\
 \label{Ccoeff}
  C&= \frac{M^2+1}{2\pi }\sum_{n=1}^{\infty}\dfrac{(-1)^n }{nn!}\,
 \mathop{\mathrm{Im}}\left(
 \,{\rm e}\,^{\frac{i\phi n}{M-i}}
 \,\dfrac{\frac{M-i}{n}\Gamma \left(\frac{M+i}{M-i}\,n\right)}{\Gamma ^2\left(\frac{i}{M-i}\,n\right)}\right).
\end{align}
The origin of the last term in $B$ is explained in appendix~\ref{anomalous-sec}.
The first of these equations solves the problem of computing the prefactor in the Wilson loop expectation value.

For small $M$,  we can approximate
\begin{equation}
 \,{\rm e}\,^{\frac{i\phi n}{M-i}}
 \,\frac{\Gamma \left(\frac{M+i}{M-i}\,n\right)}{\Gamma^2 \left(\frac{i}{M-i}\,n\right)}
\approx 
 \frac{\left(-1\right)^nn!}{2\pi }\,\tan\frac{\pi Mn}{M-i}\,. 
 \label{smallM_approx}
\end{equation}
The constant $A$ is then saturated by the $n=1$ term, which develops a pole at $M\rightarrow 0$, while the main contributions to  $B$ and $C$ come from terms in the sum with very large $n\sim 1/M$. Replacing the sums by the integrals we get:\begin{equation}\label{coeff-smallM}
A\simeq 1,\qquad 
 B\simeq \frac{M^2}{2 } \,,\qquad  \quad C\simeq \frac{M^2}{4 \pi }
 \qquad ~\left(M\rightarrow 0\right). 
\end{equation}
The constants $A$ and $B$ measure, in different ways, deviations from the Wigner distribution near the endpoint. 
We see that these deviations vanish in the $M\rightarrow 0$ limit.

Conversely, deviations from the naive strong coupling result grow with $M$ and, as we shall see, become parametrically large at $M\rightarrow \infty $. 
This limit is equivalent to the flat space limit, in which the sphere inflates to an infinite radius, corresponding to the top right hand corner of the phase diagram (fig.~\ref{PhD}) approached from below. It is here that we hope to detect signs of non-trivial phase structure,  making this regime particularly interesting. It turns out that the solution indeed develops such structure at large $M$, which we investigate in detail in the next section.

%%%%%%%%%%%%%%%%%%%%%%%%%%%%%%%%%%%%%%%%%%%%%%%%%%%%%%%%
\section{Decompactification limit}
%%%%%%%%%%%%%%%%%%%%%%%%%%%%%%%%%%%%%%%%%%%%%%%%%%%%%%%%

One of our main motivations for solving the localization matrix model at strong coupling is to study the flat space/infinite mass limit, $M R \rightarrow \infty$. The corner of the phase diagram in fig.~\ref{PhD} is an accumulation point of an infinite number of phase transitions, and even though we approach the critical point from a different direction,  we may expect to see signatures of the non-trivial phase structure in the eigenvalue density. There are three  well separated scales in the problem in the decompactification limit: the IR cutoff scale $1/R$ (equal to $1$ in the units we use), the mass scale $M$ and the symmetry breaking scale $\mu \sim \sqrt{\lambda }M$. The physics behind the phase transitions is governed by the second of these,  and we expect the non-trivial structures to occur distance $\sim M$ away from the endpoints of the eigenvalue distribution. But $M\ll\mu $ and, in spite of the fact that $M\gg 1$, the distances of order $M$ are still within the range of the scaling limit applicable near the endpoints. We thus expect that most of the interesting phenomena associated with the phase transitions are described by the solution of the Wiener-Hopf problem. We thus need to study the large-$M$ limit of the solution.

\begin{figure}[t]
  \centering
    \includegraphics[width=0.6\textwidth]{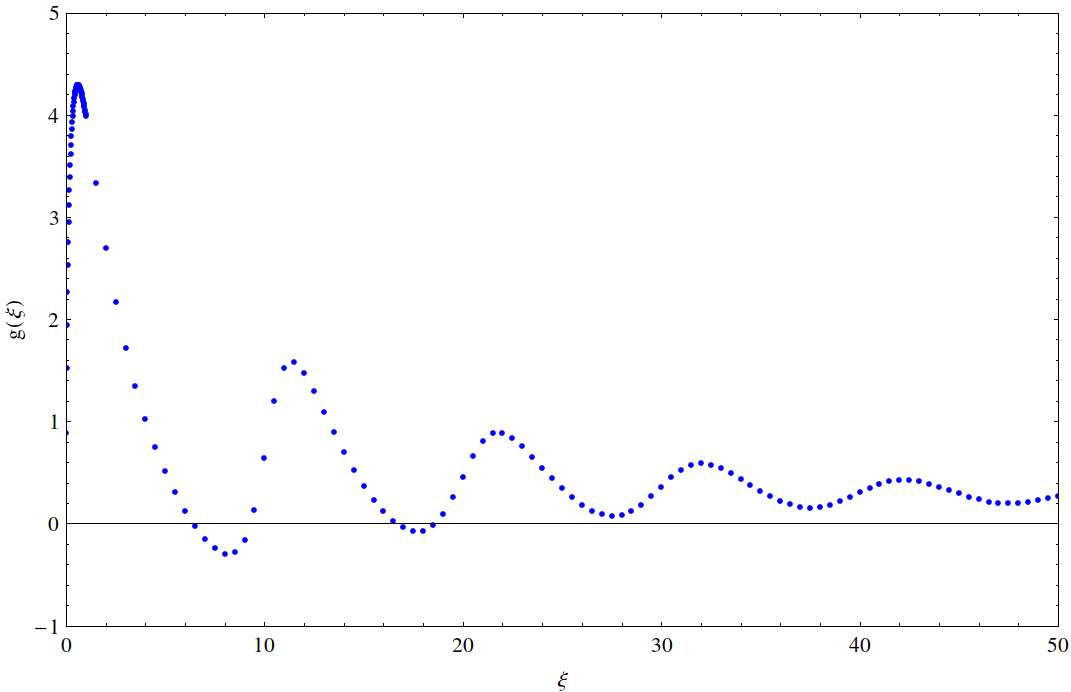}
  \caption{$g(\xi)$ evaluated numerically for $M=10$. See additional plots on page \pageref{plot_endpoint}.}
  \label{numericsM10}
\end{figure}

As $M$ grows, the shape of the scaling function $g(\xi )$ dramatically changes. At first, a smooth profile similar to that in fig.~\ref{gxi_smallM} starts to be modulated with period approximately equal to $M$. At yet larger masses, the amplitude of modulation grows, and the scaling function  develops a structure of regularly spaced peaks of diminishing amplitude (fig.~\ref{numericsM10}). The peaks become sharper and sharper with growing $M$ and in the strict $M\rightarrow \infty $  limit morph into cusps of infinite height. This is precisely the phenomenon observed in \cite{Russo:2013qaa,Russo:2013kea} where the flat space theory was studied at arbitrary coupling and the appearance of cusps was identified as a cause of the phase transitions. 

We now have an analytic expression for the scaling function which is valid at any $M$, and thus can study the large-$M$ limit rather explicitly. Moreover, the complicated expression \eqref{final-g} simplifies somewhat in this limit. As should be clear from the preceding discussion, there are two distinct regimes, of $\xi \sim 1$ and of $\xi \sim M$, which should be analyzed separately. It turns out that an additional scale arises in the UV at $\xi \sim M^2$.

%%%%%%%%%%%%%%%%%%%%%%%%%%%%%%%%%%%%%%%%%%%%%%%%%%%%%%%%
\subsection{Small $\xi$}
%%%%%%%%%%%%%%%%%%%%%%%%%%%%%%%%%%%%%%%%%%%%%%%%%%%%%%%%
In the limit $M\rightarrow \infty $ and $\omega \sim O(1)$, the sums in (\ref{final-g}) can be replaced by integrals, which leads to massive cancellations. This is not unexpected, since the solution to the Wiener-Hopf problem is designed so as to subtract the singularities of the scaling function in the upper half of the complex plane. These singularities are a series of poles (fig.~\ref{a-structureofG}), which at $M\rightarrow \infty $ collapse onto the real axis and collide with the poles in lower half plane. In effect all the singularities happen to be subtracted, leaving behind a subleading contribution without poles, whose only singularity is the square-root cut in the lower half-plane:
\begin{equation}\label{limitlargeMofg}
 \hat{g}(\omega )\simeq \frac{M}{2}\,\sqrt{\frac{i\pi }{\omega }}\,
 \frac{\,{\rm e}\,^{\frac{\omega }{\pi i}\left(\ln\frac{\omega }{2\pi i}-1\right)}}{\Gamma ^2\left(1+\frac{\omega }{2\pi i}\right)}
 \qquad \left(M\rightarrow \infty ,~\omega \sim 1\right).
\end{equation}
A derivation of this result is given in appendix~\ref{smallxi-app}. 

This limit describes the first peak in fig.~\ref{numericsM10}. The Fourier transform of (\ref{limitlargeMofg}) has more or less the same shape, shown in fig.~\ref{gxi_smallM}, as the whole scaling function at small $M$. The function grows as $M\sqrt{\xi }$, reaches a maximum and then decays as $M/(2\sqrt{\xi })$. The important difference with the small-$M$ regime is that now the scaling function is parametrically big, proportional to $M$, in contradistinction to small $M$ when the scaling function is just a small correction to the leading Gaussian result. The fact that $g(\xi )$ is the leading term now means that at large-$M$ the full density also has a peak; at smaller $M$ the peak is diluted by the growth of the square root from the Gaussian approximation, with density growing monotonically away from the endpoint.  

Notice that we cannot extract the value of the constant $C$ defined in (\ref{BC-def}), (\ref{BCfromg}) from this calculation since we tacitly assume that $\omega \gg 1/M$ and consequently cannot take the limit $\omega \rightarrow 0$. To compute $C$, we need to consider the regime $\xi \sim M\Longleftrightarrow\omega \sim  1/M$, which we do in the next section.

For the constants $A$ and $B$ defined in sec.~\ref{gen-structure-sec}, we get 
\begin{equation}
 A\simeq \frac{2\pi M}{\,{\rm e}\,^2}\,,\qquad 
 B\simeq M\qquad \left(M\rightarrow \infty \right).
\end{equation}
These results can also be obtained by applying the formulae from appendix~\ref{smallxi-app} directly to (\ref{Acoeff}), (\ref{Bcoeff}).
We see again that the deviations from the simple Gaussian model are parametrically large in the decompactification limit. 

The regime described here completely determines the Wilson loop expectation value at strong coupling, including the normalization factor. 
The Wilson loop is dual to a fundamental string in the dual supergravity background. Consequently, from the matrix model point of view, the fundamental string  probes the extreme vicinity of the endpoint of the eigenvalue distribution. The features in the eigenvalue density responsible for phase transitions are simply not visible to the fundamental string probes.

%%%%%%%%%%%%%%%%%%%%%%%%%%%%%%%%%%%%%%%%%%%%%%%%%%%%%%%%
\subsection{Oscillatory behavior}
%%%%%%%%%%%%%%%%%%%%%%%%%%%%%%%%%%%%%%%%%%%%%%%%%%%%%%%%
When $\omega \sim \mathcal{O}(1/M)$, corresponding to $\xi \sim \mathcal{O}(M)$, the sum is dominated by terms with small $n$. In this case we simply take the na\"ive large-$M$ limit of \eqref{final-g} (replacing eg. $M\pm i$  by $M$ wherever it appears). For instance, the prefactor in front of the sum in \eqref{final-g} becomes
\begin{equation}
	\frac{M^3 \omega^2}{2\sin \frac{M\omega}{2}} \,{\rm e}\,^{-\frac{i\omega M}{2}}
\end{equation}
where we have used the identity (\ref{gamma-gamma}) for the gamma functions. 

Computation of the sum involves one subtlety. It turns out that we need to keep the leading correction to the phase $\phi $:
\[
	\epsilon \equiv \pi -\frac{\phi}{M} 
	\simeq  \frac{2\ln M}{M}\,,
\]
vanishing in the $M\rightarrow \infty $ limit.  Yet this quantity leaves a finite imprint in the final answer. For the sum we get
\begin{equation}
	{\rm sum}\simeq -\frac{1}{M^2}\sum_{n=1}^\infty \left( \frac{\,{\rm e}\,^{-i \epsilon n}}{\omega - \frac{2\pi n}{M}}  +  \frac{\,{\rm e}\,^{i \epsilon n}}{\omega  + \frac{2\pi n}{M}} \right) 
	= \frac{1}{\pi M}\sum_{n=1}^\infty \frac{\frac{M\omega}{2\pi} \cos \epsilon n - i n \sin \epsilon n}{n^2-\left(\frac{M\omega}{2\pi}\right)^2} 
\end{equation}
In the \emph{cos} term we can set $\epsilon =0$ right away, because the sum converges. The \emph{sin} term does not contribute at first sight, but the sum converges slowly, and the limit $\epsilon \rightarrow 0$ does not commute with summation. Indeed,
\begin{equation}
 \lim_{\epsilon \rightarrow 0}\sum_{n=1}^{\infty }\frac{\sin\epsilon n}{n}
 =\frac{\pi }{2}\,,
\end{equation}
and not zero.  Taking this into account, we find that
\begin{equation}
  {\rm sum}=\frac{1}{M^2\omega }-\frac{\,{\rm e}\,^{\frac{iM\omega }{2}}}{2M\sin\frac{M\omega }{2}}\,.
\end{equation}

Combining  all terms together, we get for the scaling function in the large-mass limit:
\begin{equation}\label{large-mass-g}
	\hat{g}(\omega) \approx \frac{i^{3/2}\sqrt{\pi}}{2\omega \sqrt{\omega + i\epsilon}} \left[\frac{M\omega \,{\rm e}\,^{-\frac{i M \omega}{2}} }{2\sin \frac{M\omega}{2}} - 1 \right].
\end{equation}
Some remarks are in order here. In the limit we are considering, the poles of the Green's functions $G_\pm$,  shown in fig.~\ref{a-structureofG}, collapse onto the real line and merge pairwise. The unprojected part of the solution (\ref{final-g}) as a consequence has double poles in this limit. But the poles of the exact scaling function in the upper half-plane  are  eliminated by the $+$ projection, and we expect the limiting scale function to have only single poles ascending from the lower half-plane. This is exactly what happens -- the double poles get cancelled and the limiting solution has a sequence of single poles along the real line.  Their origin in the lower half-plane defines the epsilon-prescription for integrating the scaling function over frequencies.

The contour of integration in the inverse Fourier transform (\ref{FTgxi}) thus passes all the poles from above. For $\xi >0$, the contour can be closed in the lower half-plane, picking up the poles along the real axis and wrapping the branch cut along the negative imaginary axis. The scaling function becomes a sum of two terms:
\begin{equation}
	g(\xi) = h_0(\xi) + h_1(\xi)
\end{equation}

Let us consider the branch cut contribution first:
\begin{equation}
	h_1(\xi)= \frac{\sqrt{M}}{2\sqrt{\pi}} \int_0^\infty \! du\; \frac{\,{\rm e}\,^{-u \frac{\xi}{M}}}{u^{3/2}} \left(1-\frac{u }{\,{\rm e}\,^u-1}\right)
=
-\sqrt{\xi }-\frac{\sqrt{M}}{2}\,\zeta \left(\frac{1}{2}\,,1+\frac{\xi }{M}\right),
\end{equation}
where $\zeta (s,a)$ is the Hurwitz zeta function. Despite its appearance, the function $h_1(\xi )$ is always positive, starts off as a constant at $\xi =0$ and decays monotonically with $\xi$, asymptoting to the $1/\sqrt{\xi }$ tail at infinity.

The poles give
\begin{equation}
	h_0(\xi) =- i \sum_{n\neq 0}\mathop{\mathrm{Res}}\left[g(\omega) \,{\rm e}\,^{-i\omega \xi}, \; \omega = \frac{2\pi n}{M} \right]
		= \sqrt{\frac{iM}{8}} \sum_{n\neq 0} \frac{\,{\rm e}\,^{-2\pi i n \xi /M}}{\sqrt{n}}\, .
		\label{g_osc}
\end{equation}
This  function is obviously periodic with period $M$. In order to expose the periodicity, it is convenient to decompose $\xi /M$ on the integer and fractional parts:
\begin{equation}
 \left\{\frac{\xi }{M}\right\}=\frac{\xi }{M}\mathop{\mathrm{mod}}1,\qquad \left[\frac{\xi }{M}\right]=\frac{\xi }{M}-\left\{\frac{\xi }{M}\right\},
\end{equation}
keeping in mind that $h_0$ only depends on the fractional part.
The sum can again be expressed in terms of the Hurwitz zeta function:
\begin{equation}
 h_0(\xi )=\frac{\sqrt{M}}{2}\, \zeta\left(\frac{1}{2},\left\{\frac{\xi}{M}\right\}\right).
\end{equation}

\begin{figure}[t]
\begin{center}
 \subfigure[]{
   \includegraphics[height=3.9cm] {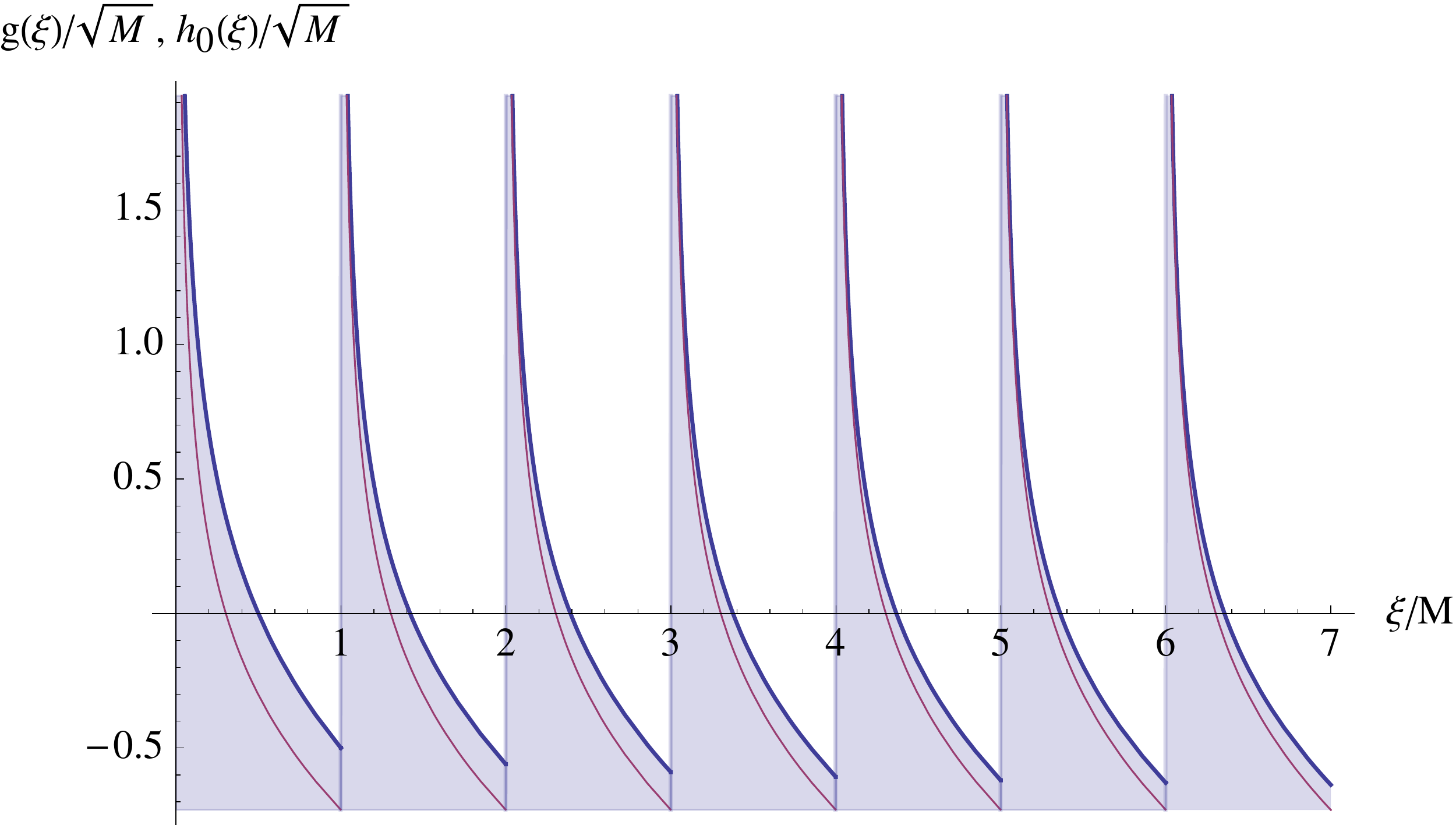}
   \label{fig-gh:subfig1}
 }
 \subfigure[]{
   \includegraphics[height=3.9cm] {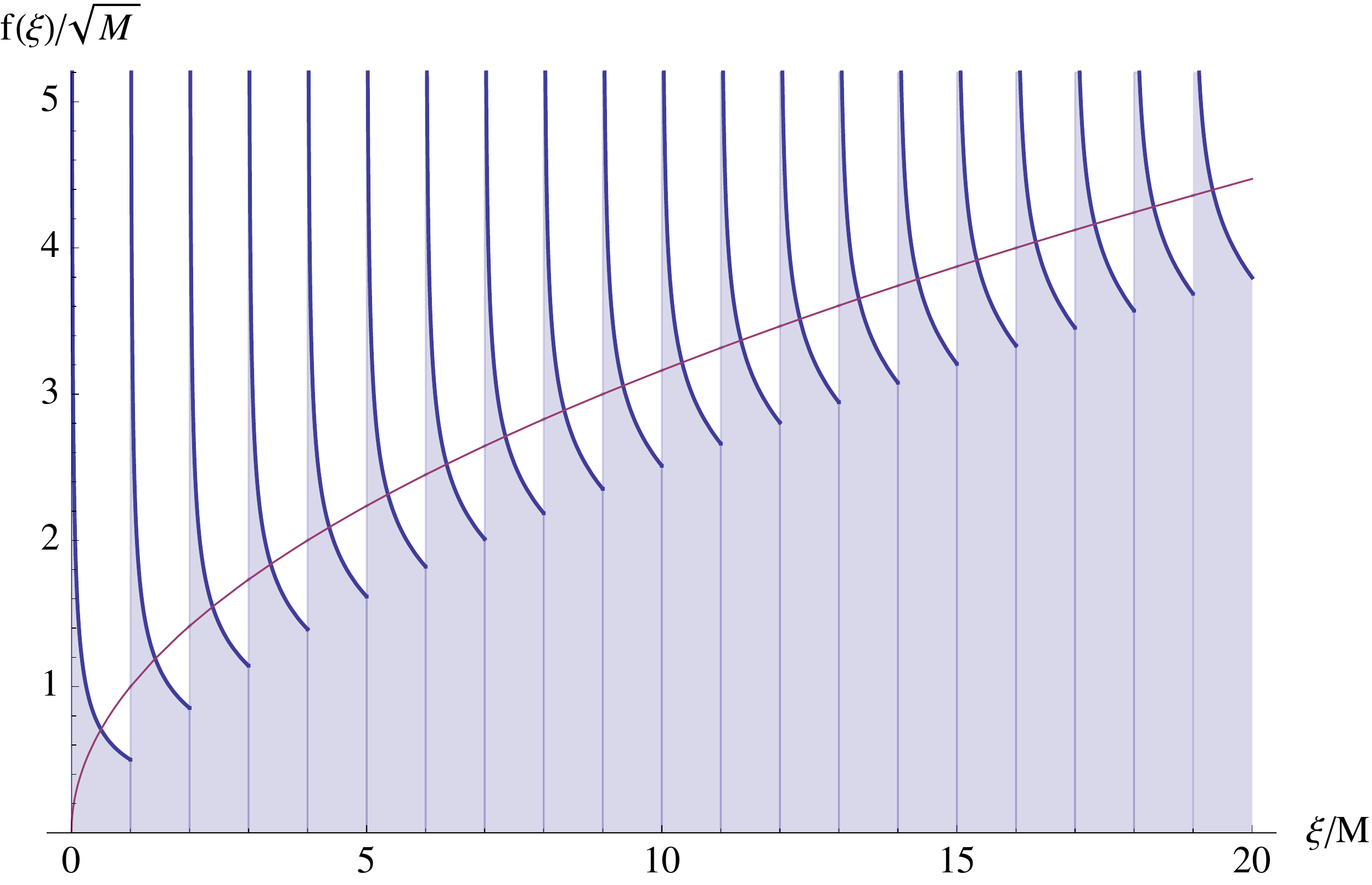}
   \label{fig-gh:subfig2}
 }
\caption{\label{fig1}\small (a) The scaling function $g(\xi )$ (upper, blue curve) and the periodic functions $h_0(\xi )$ (lower, purple curve). At larger $\xi $, $h_0(\xi )$ becomes better and better approximation to $g(\xi )$.
(b) The eigenvalue density near the endpoint  forms a comb-like structure with an infinite series of resonances on top of the leading-order square-root distribution, shown as a purple curve.}
\end{center}
\end{figure}

\begin{figure} [t] 
\begin{center}
\includegraphics[width=0.6\textwidth]{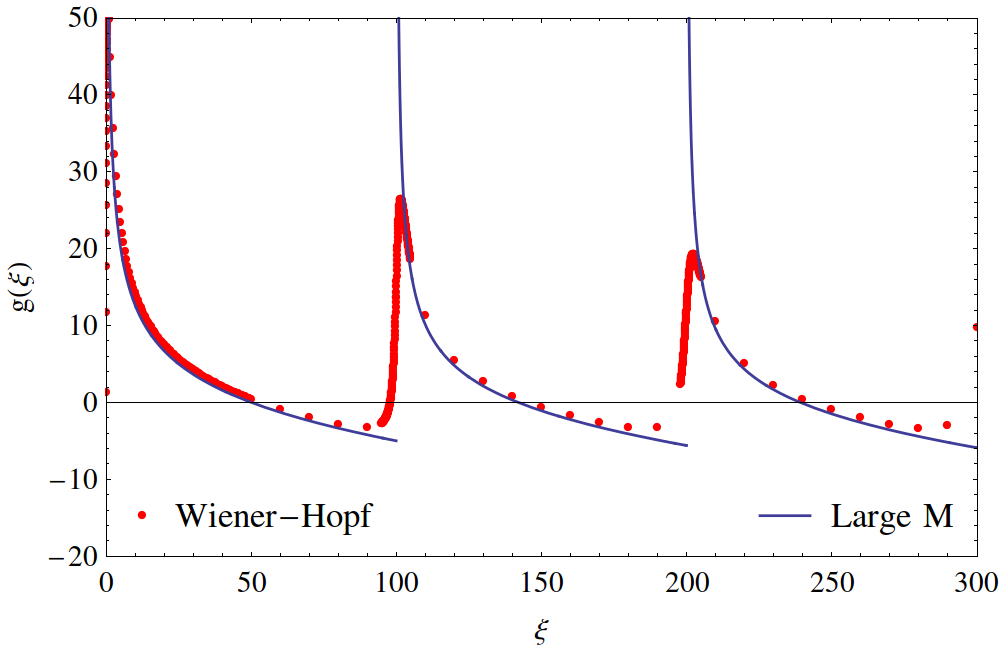} 
\end{center}
\caption{Cusp-like structure of $g(\xi)$ in the regime $\xi \sim \mathcal{O}(M)$ for large $M$. Here we compare our analytic result (\ref{gcusp_analytic}) with numerics (red) for $M=100$.} 
\label{cusp_numerics}
\end{figure}

Since $h_0(\xi )$ is periodic and $h_1(\xi )$ decreases at infinity, their sum asymptotes to $h_0(\xi )$ at $\xi \gg M$. This is illustrated in fig.~\ref{fig-gh:subfig1}.
The sum of $h_0$ and $h_1$ can be actually simplified with the help of the zeta-function identities:
\begin{equation}
 g(\xi )=\frac{\sqrt{M}}{2}\sum_{k=0}^{\left[\frac{\xi }{M}\right]}
 \frac{1}{\sqrt{\left\{\frac{\xi }{M}\right\}+k}}-\sqrt{\xi }, \label{gcusp_analytic}
\end{equation}
and, as we can see in fig.~\ref{cusp_numerics}, it agrees well with the numeric evaluation of the sum \eqref{final-g} for $M=100$.
As for the scaling form of the density, we get a particularly simple expression:
\begin{equation}
 f(\xi )=\frac{\sqrt{M}}{2}\sum_{k=0}^{\left[\frac{\xi }{M}\right]}
 \frac{1}{\sqrt{\left\{\frac{\xi }{M}\right\}+k}}\,.
\end{equation}

The function $h_0(\xi )$, and with it the scaling function $g(\xi )$, blows up as $M/(2\sqrt{\xi })$ at $\xi \rightarrow 0$. This behavior matches with the $1/\sqrt{\xi }$ tail at the upper end of  the small-$\xi $ regime, which describes the first peak of the density. The change of the endpoint exponent from $+1/2$ to $-1/2$ is characteristic of the infinite-volume limit, and is universally observed in all massive theories that can be solved on $S^4$ by localization \cite{Russo:2012ay,Russo:2013qaa,Russo:2013kea,Russo:2013sba}. But $h_0(\xi )$ is also periodic, and consequently has inverse square-root singularities in all integer points  $\xi =nM$. These are the resonances that arise due to the presence of nearly massless hypermultiplets in the spectrum. Our analysis applies to strictly infinite coupling. Varying the coupling will cause the resonances to move, resulting in phase transitions each time a full interval is traversed and a new cusp (dis)appears in the density function.

The density behaves as 
\begin{equation}
 f(\xi )\simeq \frac{M}{2\sqrt{\xi -nM}}\qquad \left(\xi \rightarrow nM^+\right),
\end{equation}
to the right of each resonance, and approaches a finite limiting value from the left. This structure is qualitatively similar to the one previously observed at finite coupling in the vicinity of the first phase transition \cite{Russo:2013kea}. But now we have an analytic solution that describes the whole resonance structure.

To move beyond the regime $\xi \sim \mathcal{O}(M)$, we should recall that the poles in $\hat{g}(\omega)$ are really located slightly off the real axis, at $\omega = \pm \frac{ 2\pi n}{M \pm i }$. This displacement means the phase in the definition \eqref{g_osc} of $h_0 (\xi)$ acquires an imaginary component $2\pi i |n| \xi/M^2$, which is unimportant for $\omega \sim \mathcal{O}(1/M)$, but for large $\xi \sim \mathcal{O}(M^2)$ causes the peaks to decay exponentially. At these large scales the oscillations in the density die out and the leading contribution is the $1/\sqrt{\xi }$ tail of the function $h_1(\xi )$:
\begin{equation}
	h_1(\xi) \simeq \frac{M}{4\sqrt{\xi}} \qquad (\xi\rightarrow \infty)
\end{equation}
This behavior determines the constant $C$ defined in (\ref{BC-def}). Comparing (\ref{large-mass-g}) at $\omega \rightarrow 0$ to  (\ref{BCfromg}) we find:
\begin{equation}
 C\simeq \frac{M}{4}\qquad \left(M\rightarrow \infty \right).
\end{equation}
This result can also be obtained directly from (\ref{Ccoeff}) by methods outlined in appendix~\ref{smallxi-app}.

To summarize, the overall picture of the eigenvalue density that emerges in the decompactification limit is as follows. 
\begin{itemize}
\renewcommand\labelitemi{--}

\item	Square root behavior at the extreme endpoint given by
\begin{equation}
	\rho (\xi)=\frac{2^{\frac{3}{2}}}{\pi \mu ^{\frac{3}{2}}}\,M\sqrt{\xi} \qquad (\xi \sim 1).
\end{equation}
The density reaches a peak while $\xi \ll M$ (corresponding to the overall peak in the small mass solution), and decays thereafter as $M/(2\sqrt{\xi})$.

\item When the density is scaled to larger $\xi \sim M$, the peak is not resolved any more and becomes a cusp, thus changing the endpoint behavior of the density from $\sqrt{\xi }$ to $1/\sqrt{\xi }$. Moreover, the density develops secondary cusps at the resonance points $\xi =nM$, and thus acquires 
 a comb-like shape with cusps separated by $M$ which are superimposed on the leading square root function. We were able to find the precise analytic form of the density in this regime: 
\begin{equation}
 	\rho(\xi )=\frac{\sqrt{2M}}{\pi \mu^{3/2}}\sum_{k=0}^{\left[\frac{\xi }{M}\right]}
 \frac{1}{\sqrt{\left\{\frac{\xi }{M}\right\}+k}}\,.
\end{equation}

\item At yet larger $\xi$, of order $M^2$, the amplitude of these resonances decays exponentially, leaving a small $1/\sqrt{\xi }$ correction to the leading-order Wigner distribution. 

\item In the bulk of the eigenvalue distribution the density is given by (\ref{corrdens}). The constant $\beta $ that controls the overall size of the correction  is determined in the next section.

\end{itemize}

\begin{figure}[t]
\centering
\subfigure[$M=10$]{%
\includegraphics[width=0.48\textwidth]{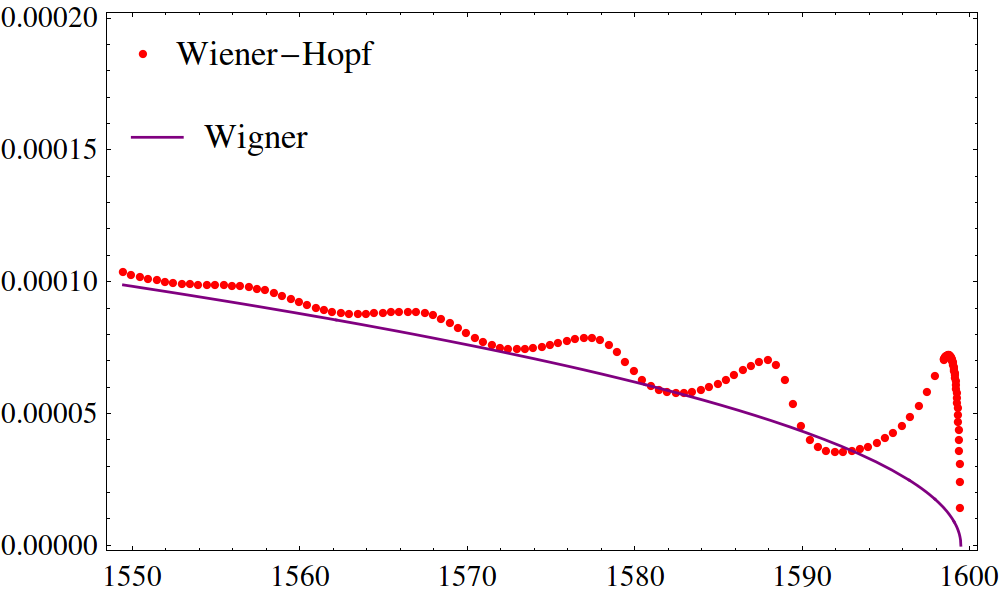}}
\,
\subfigure[$M=100$]{%
\includegraphics[width=0.48\textwidth]{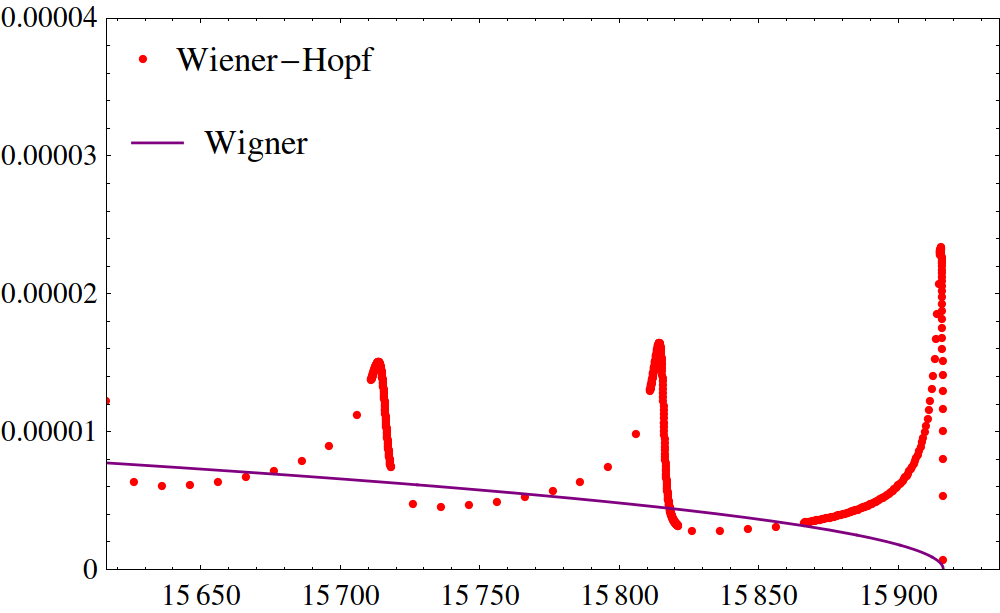}}
\caption{Endpoint behavior of the eigenvalue density, showing the leading-order semi-circle solution together with a numerical evaluation of the first-order Wiener-Hopf correction for $M=10$ and $M=100$}
\label{rhoPlots}
\end{figure}

%%%%%%%%%%%%%%%%%%%%%%%%%%%%%%%%%%%%%%%%%%%%%%%%
\section{Strong-coupling expansion}
%%%%%%%%%%%%%%%%%%%%%%%%%%%%%%%%%%%%%%%%%%%%%%%%

We now return to the question of  $1/\sqrt{\lambda }$ corrections.  As we have shown before, the functional form of the correction to the density is fixed by the saddle-point equations, but its overall normalization is not. The normalization constant can be determined by matching the bulk density to the exact scaling solution near the endpoints of the distribution.

Using the parametrization $\xi=\mu-x$, the bulk solution (\ref{corrdens}) takes the following form near the endpoint:
\begin{equation}
 \rho_{\text{bulk}}=\frac{2^{\frac{3}{2}}}{\pi \mu ^{\frac{3}{2}}} \left(\sqrt{\xi }+\frac{\pi\beta  }{4\sqrt{\xi }}\right),
\end{equation}
where we used (\ref{leading-mu}) for the endpoint position. This  has to match the asymptotic behavior of the endpoint solution at large $\xi $, for which we get combining  \eqref{fofxi-def}, (\ref{gofxi-def}) with (\ref{BC-def}):
\begin{equation}
  \rho_{\text{end}}= \frac{2^{\frac{3}{2}}}{\pi \mu ^{\frac{3}{2}}} \left(\sqrt{\xi }+g(\xi) \right)\rightarrow
  \frac{2^{\frac{3}{2}}}{\pi \mu ^{\frac{3}{2}}} \left(\sqrt{\xi }+\frac{C}{\sqrt{\xi }} \right).
\end{equation}
Comparing the two expressions  we conclude that
\begin{equation}
 \beta=\frac{4 C}{\pi}\,.
\end{equation}
For the bulk eigenvalue density we thus get
\begin{equation}\label{corrdens-bulk}
 \rho_{\rm bulk} (x)=\frac{8\pi }{\lambda \left(1+M^2\right)}\,\sqrt{\mu ^2-x^2}
 +\frac{4C}{\pi \mu \sqrt{\mu ^2-x^2}}\,,
\end{equation}
The constant $C$ is given by an infinite sum (\ref{Ccoeff}), which at small and large $M$ asymptotes to
\begin{equation}
 C\simeq \frac{M^2}{4\pi }\qquad (M\rightarrow 0),\qquad \qquad
 C\simeq \frac{M}{4}\qquad (M\rightarrow \infty ). 
\end{equation}

We can now determine the correction to the endpoint position, by imposing the normalization condition on the density:
\begin{equation}
 \int^{\mu }_{-\mu }dx \,\rho _{\rm bulk}(x)=1, 
\end{equation}
which becomes
\begin{equation}
 \dfrac{4 \pi^2  \mu ^2}{\lambda  \left(M^2+1\right)  }+\dfrac{4C}{\mu }=1,
\end{equation}
and  we find:
\begin{equation}\label{mu-2-orders}
 \mu=\dfrac{\sqrt{\lambda \left(1+M^2\right)}}{2\pi }-2C+O(\lambda^{-1/2}).
\end{equation}
The first strong-coupling correction to $\mu $ is thus simply related to the slope of the scaling function at large $\xi $. 

\begin{figure} [t] 
\begin{center}
\includegraphics[width=0.6\textwidth]{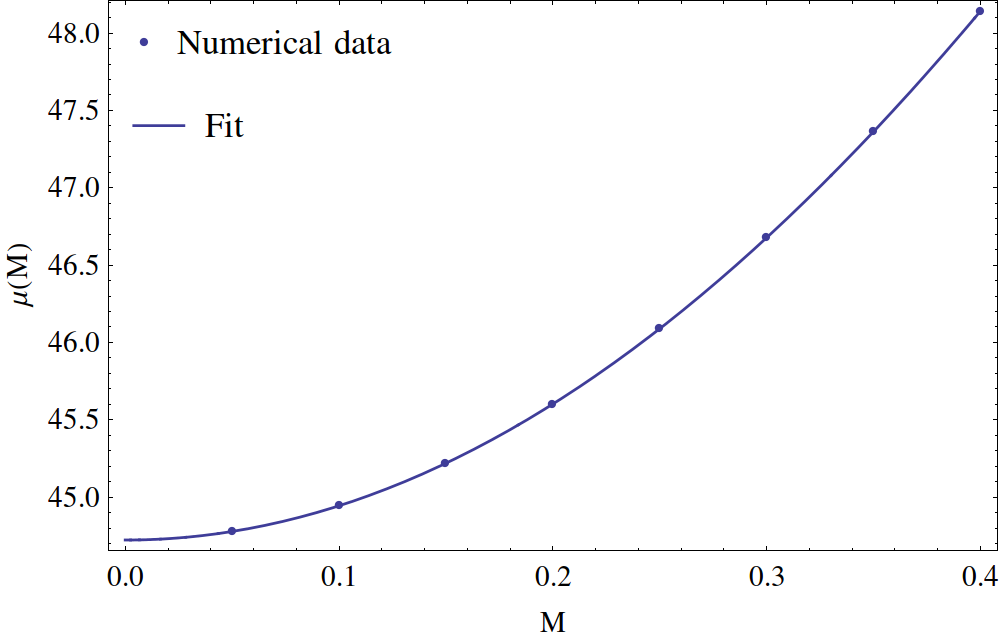} 
\end{center}
\caption{Fitting the curve $\mu(M)$ \eqref{curveMu} to the numerical data.} 
\label{fitMu}
\end{figure} 

\begin{figure} [t]
\begin{center} 
\includegraphics[width=0.49\textwidth]{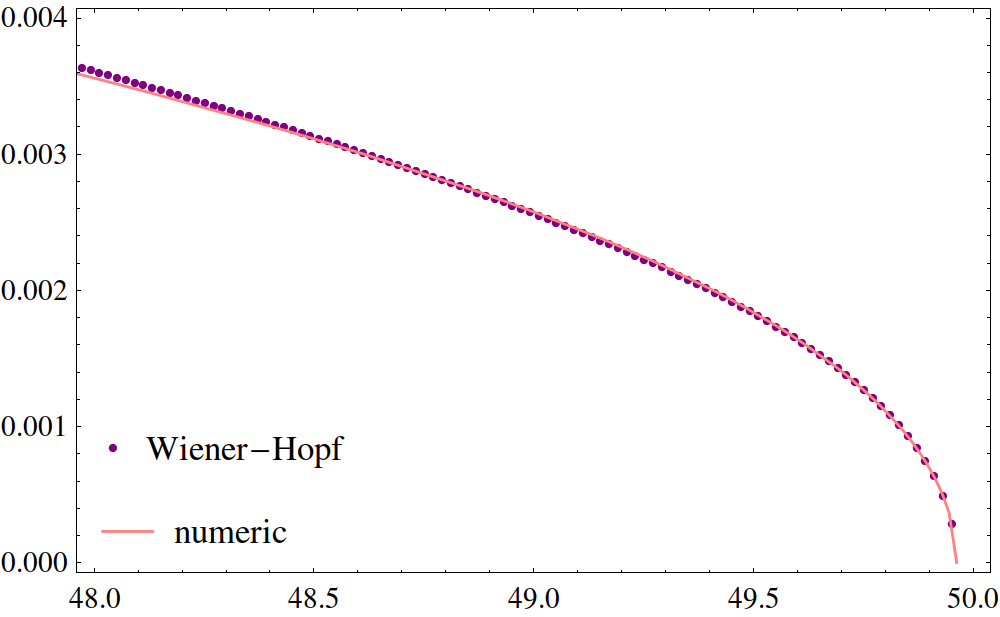} 
\includegraphics[width=0.49\textwidth]{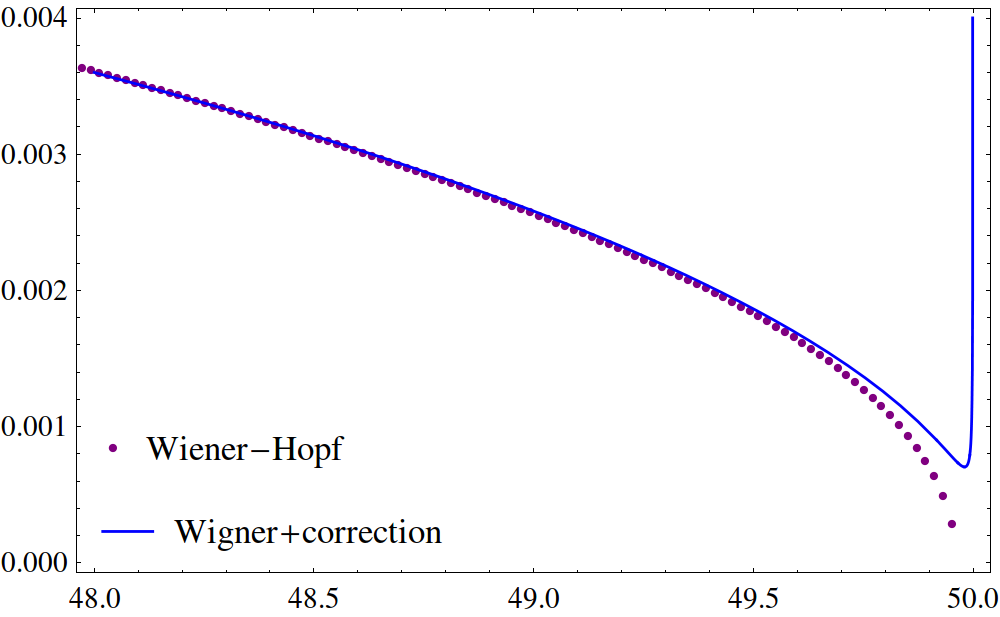} 
\end{center}
\caption{Density close to the endpoint, for $M=0.5$. 
The first plot compares the endpoint solution from Wiener-Hopf method and the direct numeric solution.
The second plot shows the matching condition of the Wiener-Hopf solution with the bulk solution \eqref{corrdens}.}
\label{plot_endpoint}
\end{figure}

This latter relation can be checked by numerically solving
the exact saddle-point equation (\ref{inteq}). The method we used becomes unstable at strong coupling unless $M$ is sufficienly small, so we restricted numerical analysis to $M<0.5$, where we can use the small-$M$ asymptotics (\ref{coeff-smallM}) for $C$. We fitted the   data points to:
\begin{equation}\label{curveMu}
 \mu(M)=a\sqrt{M^2+1}+b M^2.
\end{equation}
The results are shown in fig.~\ref{fitMu}, and are in perfect agreement with (\ref{mu-2-orders}):
\begin{center}
  \begin{tabular}{|c|c|c|c}
    \hline
    \quad & Analytic Values & Numerical Fit \\ \hline
     a & 44.7214 & 44.7221 $\pm$ 0.0004 \\ \hline
     b & -0.1591 & -0.1593 $\pm$ 0.0045 \\
    \hline
  \end{tabular}
\end{center}
In fig.~ \ref{plot_endpoint} we compare the numerical results for the density with the Wiener-Hopf solution in the endpoint region.

The correction to the endpoint position affects the normalization of the Wilson loop expectation value, through its exponential dependence on $\mu $ in  (\ref{W->A}). Taking this correction into account, we get for the Wilson loop expectation value: 
\begin{equation}\label{finalWilson}
 W(C)\simeq \sqrt{\frac{2}{\pi }}\,\frac{A\,{\rm e}\,^{-4\pi C}}{\lambda ^{\frac{3}{4}}\left(1+M^2\right)^{\frac{3}{4}}}\,\,{\rm e}\,^{\sqrt{\lambda \left(1+M^2\right)}},
\end{equation}
where the constants $A$ and $C$ are given in (\ref{Acoeff}), (\ref{Ccoeff}). In the decompactification limit, we get:
\begin{equation}\label{RsinW}
 W(C)\simeq \sqrt{\frac{8\pi} {MR}}\,\lambda ^{-\frac{3}{4}}\,{\rm e}\,^{\left(\sqrt{\lambda }-\pi \right)MR-2},
\end{equation}
where we have reinstated the dependence on the radius of the four-sphere through the rescaling $M\rightarrow MR$. 

The leading exponential term should be universal, implying that any sufficiently big Wilson loop in the $\mathcal{N}=2^*$ theory on $\mathbbm{R}^4$  should obey the perimeter law:
\begin{equation}\label{gen-W(C)}
 \ln W(C)=P(\lambda )ML\qquad (ML\gg 1),
\end{equation}
where $L$ is the length of the contour $C$. The coefficient $P(\lambda )$ governs the self-energy of an infinitely heavy quark immersed in the $\mathcal{N}=2^*$ vacuum. From (\ref{RsinW}) we find that at strong coupling the self-energy coefficient behaves as
\begin{equation}
 P(\lambda )=\frac{\sqrt{\lambda }}{2\pi }-\frac{1}{2}+O\left(\frac{1}{\sqrt{\lambda }}\right).
\end{equation}
The leading order term was computed in \cite{Buchel:2013id} and successfully compared to the area law in the Pilch-Warner geometry. The second term constitutes a prediction for the first quantum correction to the minimal area. It should be possible to compute this correction by semiclassical quantization of the string dual to the straight Wilson line.

The prefactor in (\ref{RsinW}) is in principle calculable by quantizing the string suspended on the big circle in the spherical geometry. The supergravity solution in this case is also known \cite{Bobev:2013cja}, but unfortunately only in the five-dimensional form. To compute the Wilson loop one needs to know the string action, determined by the ten-dimensional uplift of the solution.

The prefactor in the Wilson loop expectation value (\ref{RsinW}) is a  contour-dependent quantity. 
The known ten-dimensional dual of the $\mathcal{N}=2^*$ theory on $\mathbbm{R}^4$ is of little help for computing this number. 
However, the square-root scaling of the prefactor with the size of the contour may be universal and apply to any sufficiently big Wilson loop. 
The leading finite-size correction to  (\ref{gen-W(C)}) is then logarithmic with precisely known coefficient: $\delta_{\rm fin.size} \ln W(C)~=~-(1/2)\ln L$. 

%%%%%%%%%%%%%%%%%%%%%%%%%%%%%%%%%%%%%%%%%%%%%%%%
\section{Conclusions}

We studied the strong-coupling planar limit of $\mathcal{N}=2^*$ theory compactified on $S^4$. The supergravity approximation should be accurate in this regime, and since the supergravity dual of $\mathcal{N}=2^*$ SYM on $S^4$ is explicitly known\footnote{An explicit analytic solution is know at infinite radius of the sphere \cite{Pilch:2000ue}, otherwise the problem reduces to a set of ODEs that can be integrated numerically \cite{Bobev:2013cja}.}, our calculations can potentially be compared to semiclassical string theory on the dual supergravity background. The exponent in the Wilson loop expectation value (\ref{finalWilson}) should then correspond to the area of the surface bounded by the big circle of $S^4$. The prefactor can be identified with the one-loop determinant due to string fluctuations around the minimal surface. 

Interestingly, all the non-trivial features that appear at strictly infinite volume, such as consecutive phase transitions,  are visible already at the first order of the strong-coupling expansion, and it would be really interesting to explore their counterparts on the string theory side. 

As we have seen, the Wilson loop expectation value is sensitive to the immediate vicinity of the endpoint in the eigenvalue distribution, namely to distances of order one in $x$ space. In contrast, the spikes that arise in the decompactification limit are located at distances of order $O(M)$. Thus Wilson loops, or semiclassical strings, are not sensitive to the phase transitions, simply because they probe a different corner in the parameter space.

Perhaps better probes are D-branes. It is known that the eigenvalue distribution as a whole can be derived from the D-brane probe analysis \cite{Buchel:2000cn}; the question is to resolve the region distance $M$ away from the endpoints, where the density has non-trivial features in the decompactification limit. We emphasize that while the decompactification limit implies that $M\gg 1$, it always remains true that $M\ll \mu $ because of the  strong coupling. The D-brane probe analysis identifies the eigenvalue distribution with a particular locus in the dual geometry. We can now pinpoint exactly which parts of the eigenvalue distribution are responsible for the phase transitions, and we can even compute the fine structure of the eigenvalue density in this region. We can thus say that the critical behavior is associated with a particular location of the 10d space-time. It would be very interesting to understand what triggers the phase transitions in string theory.

\subsection*{Acknowledgments}

We would like to thank L.~Anderson and J.~Russo for discussions.
This work was supported by the Marie
Curie network GATIS of the European Union's FP7 Programme under REA Grant
Agreement No 317089.
The work of K.Z. was supported by the ERC advanced grant No 341222
and by the Swedish Research Council (VR) grant
2013-4329.

\appendix
\section{Exact expression for $g(\xi)$}

Here, we are going to derive a sum representation for $g(\xi)$, which is useful for studying the function numerically.

The Fourier integral \eqref{FTgxi} can be solved by applying the residue theorem, 
where we sum over the remaining poles of $ \hat{g}(\omega)$ in the lower half plane.
Notice that we also have a branch cut in the lower complex half-plane, due to the $1/\sqrt{\omega +i\epsilon }$ term in our solution \eqref{final-g}.
Our strategy is to inverse Fourier transform the branch cut and the part with poles separately. 
In the coordinate space, the final expression is then the convolution of these terms. 
% The convolution integral and the sum operator over poles commute

Let us factorize $\hat{g}(\omega)\equiv\hat{a}(\omega)\hat{b}(\omega)$, with
$\hat{a}(\omega)\equiv \dfrac{\sqrt{i \pi }}{\sqrt{\omega +i\epsilon }} $. In the coordinate space, the latter becomes:
\begin{equation}
 a(\xi)=\dfrac{1}{\sqrt{\xi}}\theta(\xi),
\end{equation}
where $\theta(\xi)$ is the Heaviside step function.

Regarding $\hat{b}(\omega)$, the only poles in the lower half plane are due to
\[
 \Gamma \left(-\dfrac{(M+i) \omega }{2 \pi }\right) \Gamma \left(\dfrac{(M-i) \omega }{2 \pi }\right),
\]
which comes from the expression for $G_+(\omega)$, \eqref{Gplusminus}. These poles are the complex conjugate of \eqref{poles}, i.e.
\begin{equation}
\bar{\omega}_n=\frac{2\pi \left(Mn-i|n|\right)}{M^2+1}\,, \quad n=\pm 1, \pm2, \ldots
\end{equation}
Applying the residue theorem to $\hat{b}(\omega)$, we obtain:
\begin{equation}
 b(\xi)=-i \sum_{n=-\infty}^\infty \text{Res}\left(\hat{b}(\omega) e^{-i \omega\xi},\bar{\omega}_n\right) \theta(\xi).
\end{equation}
Hence,
\begin{eqnarray*}
 g(\xi)&=&\int^\infty_{-\infty} \: d\eta \: a(\eta) b(\xi-\eta)\\
       &=& \int_0^{\xi} \: d\eta \: \dfrac{1}{\sqrt{\xi}} \sum_{n=-\infty}^{\infty} \text{Res}\left(\hat{b}(\omega),\bar{\omega}_n\right)e^{-i \bar{\omega}_n (\xi-\eta)}\\
\end{eqnarray*}
As the sum is convergent, the integral and the sum commute. The integral gives:
\begin{equation}
 \int_0^{\xi} \: d\eta \: \dfrac{1}{\sqrt{\eta}}e^{-i \bar{\omega}_n (\xi-\eta)}= e^{-i \bar{\omega}_n \xi} \sqrt{\dfrac{i \pi}{\bar{\omega}_n}} \text{erf}\left(\sqrt{-i \bar{\omega}_n \xi}\right)
\end{equation}
The residue term for $n=1,2,\ldots$ is explicitly:
\begin{eqnarray*}
\text{Res}\left(\hat{b}(\omega),\bar{\omega}_n\right)&=&
 \dfrac{-\left(M^2+1\right) (-1)^n }{2 \pi  n} \dfrac{\sinh ^2\left(\frac{\pi  n}{M+i}\right) }{\sinh\left(\frac{\pi  (1+i M) n}{M+i}\right)}\\
 &-&\dfrac{ \left(M^2+1\right)^2 (-1)^n \: i \pi}{M+i} 
 \dfrac{ \Gamma \left(\frac{(M+i) n}{M-i}\right) }{n! \Gamma \left(\frac{i n}{M-i}\right)^2} e^{\frac{i n \phi}{M-i}} 
 \mathcal{A}\left(\frac{2 \pi  n}{M+i}\right) 
\end{eqnarray*}
where $\mathcal{A}(\omega)$ is the sum in \eqref{final-g}.
For negative values of $n$, the residue is the negative complex conjugate of the expression above. Hence we need only the imaginary part of the positive $n$ sum. The final expression for $g(\xi)$ is thus:
\begin{eqnarray}\label{gxisolution}
 g(\xi)&=&  2\sum_{n=1}^{\infty} \Im \left[\text{Res}\left(\hat{b}(\omega),\bar{\omega}_n\right) e^{-i \bar{\omega}_n \xi} \sqrt{\dfrac{i \pi}{\bar{\omega}_n}}  
 \text{erf}\left(\sqrt{-i \bar{\omega}_n \xi}\right)\right]
\end{eqnarray}

\section{Anomalous contribution to $B$}
\label{anomalous-sec}

In computing $B$ as defined in (\ref{BCfromg}), we need to take the $\omega \rightarrow \infty $ limit of the scaling function (\ref{final-g}). The naive limit gives the first two terms in (\ref{Bcoeff}), but (\ref{final-g}) contains an infinite sum and 
one has to be careful and do the summation first, before taking $\omega \rightarrow \infty $. It turns out that the summation and taking the limit do not commute, and $B$ receives an anomalous contribution. To isolate this contribution we can divide the sum into two parts, from 1 to $N_0$ and from $N_0$ to infinity for some $N_0\gg 1$. The anomalous contribution can only come from the second part:
\begin{equation}
 \delta B_{\rm anom}=\lim_{\omega \rightarrow \infty }
 \frac{i\left(M^2+1\right)\omega }{2\pi }
 \sum_{n=N_0}^{\infty }
 \frac{1}{n}\left(
 \frac{1}{\omega -\frac{2\pi n}{M-i}}- \frac{1}{\omega +\frac{2\pi n}{M+i}}
 \right)
\end{equation}
Here we used that $n\geq N_0\gg 1$  to simplify the summand.
This expression can also be written as
\begin{equation}
 \delta B_{\rm anom}=i\left(M^2+1\right)
 \lim_{\omega \rightarrow \infty }
 \sum_{n=N_0}^{\infty }
 \left[
 \frac{1}{\left(M-i\right)\omega -2\pi n}
 +\frac{1}{\left(M+i\right)\omega +2\pi n}
 \right].
\end{equation}
The naive $\omega \rightarrow \infty $ limit would give zero, but we  need to first sum and then take the limit, and this gives a finite result:
\begin{equation}
 \delta B_{\rm anom}=\frac{i\left(M^2+1\right)}{2\pi }\,
 \lim_{\omega \rightarrow \infty }
 \ln\frac{2\pi N_0-\left(M-i\right)\omega }{2\pi N_0+\left(M+i\right)\omega }
 =-\frac{M^2+1}{\pi }\,\arctan M.
\end{equation}

\section{Large $M$ limit of scaling function}\label{smallxi-app}

Here we give the details on the derivation of the limiting expression  (\ref{limitlargeMofg}) for the scaling function from the exact one (\ref{final-g}), in the case when $M\rightarrow \infty $ and $\omega $ stays finite. We assume that $\omega $ is real throughout the derivation. 

We start by examining the infinite sums appearing in (\ref{final-g}):
\begin{equation}\label{mathcal-A}
 \mathcal{A}_\pm=\sum_{n=1}^{\infty }\frac{a_\pm\left(\frac{n}{M\pm i}\right)}{\omega \pm\frac{2\pi n}{M\pm i}}\,,
\end{equation}
where
\begin{equation}
 a_\pm(x)=\frac{\,{\rm e}\,^{\mp i\left[\phi -(M\pm i)\pi \right]x}}{(M\pm i)^2x^2}\,\,
 \frac{\Gamma \left((M\mp i)x\right)}{\Gamma \left((M\pm i)x\right)\Gamma ^2\left(\mp ix\right)}\,.
\end{equation}
In terms of these sums,
\begin{eqnarray}\label{final-g-1}
 \hat{g}(\omega )&=&\frac{i^{\frac{3}{2}}\sqrt{\pi }}{2\omega^{\frac{3}{2}} }
 \left[
 \frac{M^2\sinh^2\frac{\omega }{2}-\sin^2\frac{M\omega }{2}}{\sinh^2\frac{\omega }{2}+\sin^2\frac{M\omega }{2}}
\right.
\nonumber \\
 &&\left.
 +\left(M^2+1\right)^2\omega \,{\rm e}\,^{- \frac{i\phi \omega}{2\pi } }\,
 \frac{\Gamma \left(\frac{M- i}{2\pi }\,\omega \right)
 \Gamma \left(-\frac{M+ i}{2\pi }\,\omega \right)}{\Gamma ^2\left(-\frac{i\omega }{2\pi }\right)}
\left(\mathcal{A}_-+\mathcal{A}_+\right) \right]. 
\end{eqnarray}

Since $a_\pm(x)$ has a finite limiting value at zero:
\begin{equation}
 a_\pm(0)=-\frac{1}{M^2+1}\,,
\end{equation}
the sums (\ref{mathcal-A}) appear linearly divergent if $M$ is sent to infinity independently in each term. The main contribution consequently comes from very large $n\sim M$, because then $a_\pm(x)$  become slowly varying functions of their argument, namely
\begin{equation}
 a_\pm(x)\stackrel{M \rightarrow \infty }{\simeq }
 \frac{\,{\rm e}\,^{\mp 2ix\left(\ln(\mp ix)-1\right)}}{M^2x^2\Gamma ^2\left(\mp ix\right)}\,,
\end{equation}
assuming $x>0$.
We are not going to use these approximate expressions, because of the necessity to keep the next-to-leading order accuracy.  It will suffice to know that the exact $a_+(z)$ is an analytic function in the upper half plane, decreases as $1/z$ in its domain of analyticity, and satisfies  the  following functional identity:
\begin{equation}\label{identitya+a-}
 (M+i)a_+(-x)=(M-i)a_-(x)\,\frac{\sin\pi (M+i)x}{\sin\pi (M-i)x}\,\,{\rm e}\,^{2\pi x}.
\end{equation}

Normally, the sum of $f(n/M)$, where $M$ is a big paramater, is well approximated by the intergal with the help of the Euler-Maclaurin formula, but here we need to be more careful. The summands in $\mathcal{A}_\pm$ have poles at $2\pi n/(M\pm i)=\omega $ that collapse onto the contour of integration in the $M\rightarrow \infty $ limit. In the vicinity of the poles the summand is not a slowly varying function of $n/M$, and the summation has to be performed exactly. As a result a more general formula applies:
\begin{equation}\label{curious-old}
 \mathcal{A}_\pm\simeq (M\pm i)\strokedint_0^\infty \frac{dx\,a_\pm(x)}{\omega \pm 2\pi x}-\frac{a_\pm(0)}{2\omega }
 +\frac{M\pm i}{2}\,\theta (\mp\omega )a_\pm\left(\mp\frac{\omega }{2\pi }\right)\cot\frac{(M\pm i)\omega }{2}\,.
\end{equation}
The second term is the Euler-Maclaurin correction, which we need to make this formula correct throughout the next-to-leading order. The last term is the result of summation around $n\sim M\omega /2\pi $. We excluded this region from the integral by the principal-value prescription, to avoid double-counting. This formula can be viewed as a contour-deformation prescription that takes into account the discreteness of the sum in the residue term. The formula can be brought to the form
\begin{equation}\label{curious}
 \mathcal{A}_\pm\simeq (M\pm i)\int_0^\infty \frac{dx\,a_\pm(x)}{\omega \pm 2\pi x +i\epsilon }-\frac{a_\pm(0)}{2\omega }
 +\frac{M\pm i}{2}\,\theta (\mp\omega )a_\pm\left(\mp\frac{\omega }{2\pi }\right)\frac{\,{\rm e}\,^{\frac{i(M\pm i)\omega }{2}}}{\sin\frac{M\pm i}{2}\,\omega }\,,
\end{equation}
that facilitates rotation of the contour of integration into the domain of analyticity of the integrand.

Substitution of (\ref{curious}) into (\ref{final-g-1}) leads to massive cancellations. Let us first concentrate on the integral terms. Using the identity (\ref{identitya+a-}) we can transform their sum as
\begin{eqnarray}\label{int-terms}
 &&(M+ i)\int_0^\infty \frac{dx\,a_+(x)}{\omega + 2\pi x+i\epsilon }+
 (M- i)\int_0^\infty \frac{dx\,a_-(x)}{\omega - 2\pi x+i\epsilon }
\nonumber \\
&&
 =(M+i)\int_{-\infty }^\infty \frac{dx\,a_+(x)}{\omega + 2\pi x+i\epsilon }
 \nonumber \\
&&
 +(M- i)\int_0^\infty \frac{dx\,a_-(x)}{\omega - 2\pi x+i\epsilon }
 \left(1-\frac{\sin(\pi Mx+ix)}{\sin(\pi Mx-ix)}\,\,{\rm e}\,^{2\pi x}\right).
\end{eqnarray}
We are going to argue that both terms are negligible in the large-$M$ limit. The first integral actually vanishes identically, which follows from the contour argument since the intergand has no singularities in the upper half plane.
The second integral  contains a rapidly oscillating function that depends on the slow variable $x$ and the fast variable $\pi Mx$. The dependence on the fast variable is periodic, and integration over $x$ goes through many periods of oscillations before the slow dependence on $x$ can substantially alter the integrand. In any integral of this type, the integrand $\mathcal{F}(\pi Mx,x)$ can  be replaced by its average:
\begin{equation}
\int_{}^{}dx\,\mathcal{F}(\pi Mx,x)\simeq \int_{}^{}dx\, \left\langle \mathcal{F}(\Omega ,x)\right\rangle,\qquad 
 \left\langle \mathcal{F}(\Omega ,x)\right\rangle\equiv 
 \int_{0}^{2\pi }\frac{d\Omega }{2\pi }\,\,\mathcal{F}(\Omega ,x).
\end{equation}
It is easy to show that
\[
\left\langle 1-\frac{\sin(\Omega +ix)}{\sin(\Omega -ix)}\,\,{\rm e}\,^{2\pi x} \right\rangle=0
\]
for $x>0$. We can therefore drop the last integral in (\ref{int-terms}) too. 

A relatively long but straightforward calculation shows that the residue term in (\ref{curious}) cancels with the first term in the square brackets in (\ref{final-g-1}) with the requisite, $O(M^0)$ accuracy. Thus only the Euler-Maclaurin term contributes and we are left with
\begin{equation}
 \hat{g}(\omega )=\frac{i^{\frac{3}{2}}\sqrt{\pi }M^2}{2\omega^{\frac{3}{2}} }\,
 \,{\rm e}\,^{- \frac{i\phi \omega}{2\pi } }\,
 \frac{\Gamma \left(\frac{M- i}{2\pi }\,\omega \right)
 \Gamma \left(-\frac{M+ i}{2\pi }\,\omega \right)}{\Gamma ^2\left(-\frac{i\omega }{2\pi }\right)}\,.
\end{equation}
Using the basic gamma-function identity (\ref{gamma-gamma}), and the Stirling formula we get at large $M$:
\begin{equation}
 \hat{g}(\omega )\simeq \frac{i^{\frac{3}{2}}\sqrt{\pi }M}{4\sqrt{\omega }}\,\,
 \frac{\,{\rm e}\,^{\frac{\omega }{\pi i}\left(\ln\frac{\omega }{2\pi i}-1\right)}}{\Gamma ^2\left(1+\frac{\omega }{2\pi i}\right)}\,\,
 \frac{\,{\rm e}\,^{-\frac{iM\omega }{2}+\frac{|\omega |}{2}}}{\sin\left(\frac{M\omega }{2}+\frac{i|\omega |}{2}\right)}\,.
\end{equation}
 The last factor still depends on $M$ through the periodic dependence on the fast variable $M\omega /2$. The Fourier transform back to the $\xi $-representation will effectively average over rapid oscillations, so the last factor can be replaced with
\begin{equation}
 \left\langle \frac{\,{\rm e}\,^{-i\Omega +\frac{|\omega |}{2}}}{\sin\left(\Omega +\frac{i|\omega |}{2}\right)}\right\rangle=\frac{2}{i}\,,
\end{equation}
which gives the final result (\ref{limitlargeMofg}) quoted in the main text.

\bibliographystyle{nb}
%\bibliography{refs}

\end{document}